\documentclass[aps,pre,twocolumn,showpacs,groupedaddress]{revtex4}
\usepackage{graphicx}
\usepackage{dcolumn}
\usepackage{bm}
\usepackage{amssymb}
\usepackage{color}

\begin{document}


\title{Shape deformation of lipid membranes by banana-shaped protein rods: Comparison with isotropic inclusions
 and membrane rupture} 

\author{Hiroshi Noguchi}
\email[]{noguchi@issp.u-tokyo.ac.jp}
\affiliation{
Institute for Solid State Physics, University of Tokyo,
 Kashiwa, Chiba 277-8581, Japan}

\date{\today}

\begin{abstract}
The assembly of curved protein rods on fluid membranes is studied using implicit-solvent 
meshless membrane simulations. 
As the rod curvature increases, the rods on a membrane tube assemble
along the azimuthal direction first and subsequently along the longitudinal direction.
Here, we show that both transition curvatures decrease with increasing rod stiffness.
For comparison, curvature-inducing isotropic inclusions are also simulated.
When the isotropic inclusions have the same bending rigidity as the other membrane regions,
the inclusions are uniformly distributed on the membrane tubes and vesicles
even for large spontaneous curvature of the inclusions.
However, the isotropic inclusions with much larger bending rigidity
induce shape deformation and are concentrated on the region of a preferred curvature.
For  high rod density, high rod stiffness, and/or low line tension of the membrane edge,
the rod assembly induces vesicle rupture, resulting in the formation of a high-genus vesicle. 
A gradual change in the curvature suppresses this rupture.
Hence, large stress, compared to the edge tension, induced by the rod assembly  is the key factor 
determining rupture.
For rod curvature with the opposite sign to the vesicle curvature, 
membrane rupture induces inversion
of the membrane, leading to division into multiple vesicles as well as
formation of a high-genus vesicle. 
\end{abstract}



\maketitle

\section{Introduction}

In living cells, biomembranes often dynamically change their shapes to carry out their biological functions.
Membrane budding, fusion, and fission occur during endo/exocytosis and vesicle transports.
Cell organelles have specific shapes depending on their functions.
Various types of proteins participate in regulation of these dynamic and static membrane shapes \cite{mcma05,shib09,drin10,baum11,joha14,mcma15}.
These proteins mainly control local membrane shapes in two ways: hydrophobic insertions (wedging) and scaffolding.
In the former mechanism, a part of the protein, such as an amphipathic $\alpha$-helix, is inserted into the lipid bilayer membrane.
In the latter mechanism, the protein domain has a strong affinity for the lipid polar head groups
and adsorbs onto the lipid membrane.
A BAR (Bin/Amphiphysin/Rvs) domain, which consists of a banana-shaped dimer,
mainly bends the membrane along the domain axis via scaffolding~\cite{itoh06,masu10,zhao11,mim12a,simu15a}.
Some of the BAR superfamily proteins, such as N-BAR proteins, also have hydrophobic insertions.
Experimentally, the formation of membrane tubes and curvature-sensing by various types of BAR superfamily proteins have been observed \cite{itoh06,masu10,zhao11,mim12a,simu15a,pete04,matt07,fros08,wang09,zhu12,tana13,shi15,prev15,isas15,adam15}.
Dysfunctional BAR proteins are considered to be implicated in neurodegenerative, cardiovascular, and neoplastic diseases.
Thus, it is important to understand the mechanism by which membrane shaping is regulated by proteins, 
not only from a basic science perspective but also for medical applications.

Homogeneous lipid membranes in a fluid phase are laterally isotropic and have zero spontaneous curvature.
A local non-zero isotropic spontaneous curvature can be induced by adhesion of spherical colloids and 
polymer-anchoring, as well as by transmembrane and other proteins \cite{lipo13}.
Here, we call these objects that induce isotropic  spontaneous curvature an isotropic inclusion.
Their assembly into  preferred curvature regions \cite{gozd06,phil09,gree11,nogu12a,aimo14}
and membrane-mediated interactions between the colloids \cite{reyn07,auth09,sari12}  have been previously explored.

In contrast, BAR domains, which are banana-shaped, generate an anisotropic curvature.
Amphipathic $\alpha$-helices can also yield an anisotropic curvature \cite{gome16}.
Recently, the anisotropic nature of curvature has received increasing attention theoretically.
The classical Canham--Helfrich curvature free energy \cite{canh70,helf73} 
has been extended to anisotropic curvatures \cite{four96,kaba11,igli06}.
To simplify the interactions, the protein and membrane underneath it have been often modeled together as an undeformable object with a fixed curved shape
such as a point-like object with an anisotropic curvature \cite{domm99,domm02} and a bent elliptical surface \cite{schw15}.
Furthermore, it has also been clarified that two undeformable  parallel rods have an attractive interaction
but the interaction is repulsive for a perpendicular orientation.

Atomic and coarse-grained molecular simulations \cite{bloo06,arkh08,khel09,yu13,simu13,simu15} 
have been employed to investigate molecular-scale interactions between BAR proteins and lipids. 
The scaffold formation~\cite{yu13} and linear assembly~\cite{simu13} of BAR domains have been demonstrated.
To investigate large-scale membrane deformations,
a dynamically triangulated membrane model \cite{rama12,rama13} 
and meshless membrane models \cite{nogu14,nogu15b,nogu16,ayto09} have been employed;
consequently, various (meta)stable vesicle shapes~\cite{rama12,rama13,nogu14,nogu15b}
and the tubule formation dynamics \cite{nogu16} have been reported. 
Using meshless membrane and molecular simulations, vesicle rupture into high-genus vesicles has also been investigated \cite{ayto09,simu13a}.
The high-genus vesicles obtained in this way resemble electron microscopic images of high-genus liposomes induced by N-BAR proteins well \cite{ayto09,simu13a}. 
Despite these numerous advancements, many questions related to the coupling between membrane shape deformation and the assembly of the protein rods remain.

In this paper, we focus on three questions: (i) How does protein elasticity modify protein assembly?
(ii) How is rod assembly different from the isotropic inclusions?
(iii) How is membrane rupture induced by protein rods?
In the previous rupture simulations~\cite{ayto09,simu13a}, the effects of bending rigidity and rod curvature have been investigated 
but the other mechanical properties were not varied.
We show here that the line tension of the membrane edge and the annealing speed, 
as well as the rod stiffness and density, are important parameters that determine the condition of membrane rupture.

In Sec.~\ref{sec:method}, the simulation model and method are described.
We simulate membrane tubes and vesicles
using an implicit-solvent meshless membrane model~\cite{nogu09,nogu06,shib11,nogu13,nogu14,nogu15b,nogu16}.
A banana-shaped protein rod is  assumed to be strongly adsorbed onto the membrane
and the protein and membrane region below it are modeled as a linear string of particles with a bending stiffness and preferred curvature.
In order to investigate the membrane-mediated interactions,
no direct attractive interaction is considered between the rods.

In Sec.~\ref{sec:iso}, the coupling between the assembly of the isotropic inclusions and shape deformation of membrane tubes and vesicles 
are presented.
In Sec.~\ref{sec:rodtube}, the assembly of the protein rods in the membrane tubes is shown.
The dependence on the rod stiffness is investigated and the results are compared with those of the isotropic inclusions.
In Sec.~\ref{sec:rup}, the vesicle rupture into high-genus vesicles and vesicle division are presented.
The summary is given in Sec.~\ref{sec:sum}.

\section{Simulation Model and Method}\label{sec:method}

\subsection{Membrane Model}

We employ a spin meshless membrane model \cite{shib11,nogu14,nogu15b,nogu16}.
The details of this meshless membrane model are described in Ref.~\onlinecite{shib11}.
The position and orientational vectors of the $i$-th particle are ${\bf r}_{i}$ and ${\bf u}_i$, respectively.
The membrane particles interact with each other via a potential,
\begin{eqnarray}
\frac{U}{k_{\rm B}T} &=\ \ & \hspace{1cm} \sum_{i<j} U_{\rm {rep}}(r_{i,j}) \label{eq:U_all}
               +\varepsilon \sum_{i} U_{\rm {att}}(\rho_i)  \\ \nonumber
&\ \ +& \ \ \frac{k_{\rm{tilt}}}{2} \sum_{i<j} \bigg[ 
( {\bf u}_{i}\cdot \hat{\bf r}_{i,j})^2
 + ({\bf u}_{j}\cdot \hat{\bf r}_{i,j})^2  \bigg] w_{\rm {cv}}(r_{i,j}) \\ \nonumber
&\ \ +&  \frac{k_{\rm {bend}}}{2} \sum_{i<j}  \bigg({\bf u}_{i} - {\bf u}_{j} - C_{\rm {bd}} \hat{\bf r}_{i,j} \bigg)^2 w_{\rm {cv}}(r_{i,j}),
\end{eqnarray} 
where 
${\bf r}_{i,j}={\bf r}_{i}-{\bf r}_j$, $r_{i,j}=|{\bf r}_{i,j}|$,
and $\hat{\bf r}_{i,j}={\bf r}_{i,j}/r_{i,j}$.
Each particle has an excluded volume with a diameter $\sigma$ that results from the repulsive potential,
$U_{\rm {rep}}(r)=\exp[-20(r/\sigma-1)]$,
with a cutoff at $r=2.4\sigma$.

The second term in Eq.~(\ref{eq:U_all}) represents the attractive interaction between particles.
An attractive multibody potential $U_{\rm {att}}(\rho_i)$ is 
employed to allow the formation of a fluid membrane over wide parameter ranges.
The potential $U_{\rm {att}}(\rho_i)$ is given by
\begin{eqnarray} \label{eq:U_att}
U_{\rm {att}}(\rho_i) = 0.25\ln[1+\exp\{-4(\rho_i-\rho^*)\}]- C.
\end{eqnarray} 
Here,  $\rho_i= \sum_{j \ne i} f_{\rm {cut}}(r_{i,j})$ and $C= 0.25\ln\{1+\exp(4\rho^*)\}$,
where $f_{\rm {cut}}(r)$ is a $C^{\infty}$ cutoff function \cite{nogu06}
\begin{equation} \label{eq:cutoff}
f_{\rm {cut}}(r)=\left\{ 
\begin{array}{ll}
\exp\{A(1+\frac{1}{(r/r_{\rm {cut}})^n -1})\}
& (r < r_{\rm {cut}}) \\
0  & (r \ge r_{\rm {cut}}) 
\end{array}
\right. ,
\end{equation}
with $A=\ln(2) \{(r_{\rm {cut}}/r_{\rm {att}})^n-1\}$,
$r_{\rm {att}}= 1.9\sigma$  $(f_{\rm {cut}}(r_{\rm {att}})=0.5)$, 
and the cutoff radius $r_{\rm {cut}}=2.4\sigma$.
Here, $n=6$ is employed, as described in Ref.~\cite{nogu14}, instead of $n=12$, as described in Ref.~\cite{shib11},
to use a less steep function.
The density $\rho^*=7$ in $U_{\rm {att}}(\rho_i)$ is the characteristic density at which
the attraction is smoothly truncated.
For $\rho_i < \rho^*-1$,
$U_{\rm {att}}(\rho_i)$ acts as a pairwise attractive potential, 
while it approaches a constant value for $\rho_i > \rho^*+1$.

The third and fourth terms in Eq.~(\ref{eq:U_all}) are
discretized versions of the
tilt and bending potentials, respectively.
A smoothly truncated Gaussian function~\cite{nogu06} 
is employed as the weight function 
\begin{equation} \label{eq:wcv}
w_{\rm {cv}}(r)=\left\{ 
\begin{array}{ll}
\exp (\frac{(r/r_{\rm {ga}})^2}{(r/r_{\rm {cc}})^n -1})
& (r < r_{\rm {cc}}) \\
0  & (r \ge r_{\rm {cc}}) 
\end{array}
\right. ,
\end{equation}
where  $n=4$, $r_{\rm {ga}}=1.5\sigma$, and $r_{\rm {cc}}=3\sigma$.

The membranes are in a fluid phase over a wide range of these parameters,
and the properties of the fluid membrane can be widely varied.
The spontaneous curvature $C_0$ of the membrane is 
given by $C_0\sigma= C_{\rm {bd}}/2$ \cite{shib11}.
For the membrane particles, which do not consist of proteins,
 $k_{\rm{tilt}}=k_{\rm {bend}}=k_{\rm {mb}}=10$ and $C_{\rm {bd}}=0$ are used.
The bending rigidity $\kappa$ and edge tension $\Gamma$ can be controlled by $k_{\rm {mb}}$ and $\varepsilon$, respectively.
$\kappa$ is a linear function of $k_{\rm {mb}}$: $\kappa/k_{\rm B}T= 1.77k_{\rm {mb}} -2.5$ at $\varepsilon/k_{\rm B}T=5$,
while $\kappa$ increases only $10$\% from $\varepsilon/k_{\rm B}T=3.5$ to $8$.
$\Gamma$ is a monotonic increasing function of $\varepsilon$:
$\Gamma \sigma/k_{\rm B}T=4.4$, $5.7$, and $6.8$ at $\varepsilon/k_{\rm B}T=3.5$, $5$, and $8$, respectively.
We fix $\varepsilon/k_{\rm B}T=5$ except for the vesicle-rupture simulations.

\subsection{Protein Model}

A laterally isotropic membrane inclusion is modeled 
as a membrane particle with $k_{\rm r}$ times larger bending rigidity and isotropic spontaneous curvature $C_{\rm {iso}}$.
For the neighbor pair of inclusions, $k_{\rm{tilt}}=k_{\rm {bend}}=k_{\rm r}k_{\rm {mb}}$
and $C_{\rm {bd}}=2C_{\rm {iso}}\sigma$ are employed in Eq.~(\ref{eq:U_all}).
For the pair of an inclusion and a membrane particle, averaged values are used as 
$k_{\rm{tilt}}=k_{\rm {bend}}=k_{\rm {mb}}(k_{\rm r}+1)/2$
and $C_{\rm {bd}}=C_{\rm {iso}}\sigma$.

Note that, if the same values are used for a pair of an inclusion and a membrane particle,
 an additional attraction between the inclusions is induced by depletion,
since the inclusion assembly reduces the area of the large bending rigidity.
A similar attraction has previously been obtained for the binding sites of two membranes, when the membranes around the binding sites are hardened~\cite{nogu13}.

The protein rod is modeled as a linear chain of $N_{\rm {sg}}$ membrane particles.
We use $N_{\rm {sg}}=10$ and a rod length of $r_{\rm {rod}}=10\sigma$, which corresponds to the typical aspect ratio of BAR domains.
The BAR domain width is approximately $2$ nm, and its length ranges from $13$ to $27$ nm \cite{masu10}.
The protein particles in each protein rod
 are connected by a bond potential $U_{\rm {rbond}}/k_{\rm B}T = (k_{\rm {rbond}}/2\sigma^2)(r_{i+1,i}-l_{\rm rod})^2$.
The bending potential is given by $U_{\rm {rbend}}/k_{\rm B}T = (k_{\rm {rbend}}/2)(\hat{\bf r}_{i+1,i}\cdot\hat{\bf r}_{i,i-1}- C_{\rm r})^2$,
where $C_{\rm r}=1- (C_{\rm {rod}}l_{\rm rod})^2/2$.
We use $k_{\rm {rbond}}=40$, $k_{\rm {rbend}}=4000$, and $l_{\rm rod}=1.15\sigma$.
The membrane potential parameters between neighboring protein particles in each rod are modified as
$k_{\rm{tilt}}=k_{\rm {bend}}=k_{\rm r}k_{\rm {mb}}$ and $C_{\rm {bd}}=2C_{\rm {rod}}\sigma$
in order to ensure bending of the rod along the normal to the membrane surface.

\subsection{Simulation Method} 

The motion of the particle position ${\bf r}_{i}$ and 
the orientation ${\bf u}_{i}$ are given by underdamped Langevin equations:
\begin{eqnarray} \label{eq:lan1}
  \frac{d {\bf r}_{i}}{dt} &=& {\bf v}_{i}, \ \  \frac{d {\bf u}_{i}}{dt} = {\boldsymbol \omega}_{i}, \\
m \frac{d {\bf v}_{i}}{dt} &=&
 - \zeta_{0} {\bf v}_{i} + {\bf g}^{0}_{i}(t)
 + {{\bf f}_i}, \label{eq:lan2}  \\ \label{eq:lan3}
I \frac{d {\boldsymbol \omega}_{i}}{dt} &=&
 - \zeta_{\rm r} {\boldsymbol \omega}_i + ({\bf g}^{\rm r}_{i}(t)
 + {{\bf f}_i}^{\rm r})^{\perp} + \lambda_{\rm L} {\bf u}_{i},
\end{eqnarray}
where $m$ and $I$ are the mass and moment of inertia of the particle, respectively.
The forces are given by ${{\bf f}_i}= - \partial U/\partial {\bf r}_{i}$
and ${{\bf f}_i}^{\rm r}= - \partial U/\partial {\bf u}_{i}$ with 
the perpendicular component ${\bf a}^{\perp} ={\bf a}- ({\bf a}\cdot{\bf u}_{i}) {\bf u}_{i}$
and a Lagrange multiplier $\lambda_{\rm L}$ to keep ${\bf u}_{i}^2=1$.
According to  the fluctuation--dissipation theorem,
the friction coefficients $\zeta_{0}$ and $\zeta_{\rm r}$ and 
the Gaussian white noises ${\bf g}^{0}_{i}(t)$ and ${\bf g}^{\rm r}_{i}(t)$
obey the following relations of their averages and variances:
\begin{eqnarray}
\langle g^{\beta_1}_{i,\alpha_1}(t) \rangle  &=& 0, \\ \nonumber
\langle g^{\beta_1}_{i,\alpha_1}(t_1) g^{\beta_2}_{j,\alpha_2}(t_2)\rangle  &=&
         2 k_{\rm B}T \zeta_{\beta_1} \delta _{ij} \delta _{\alpha_1 \alpha_2} \delta _{\beta_1 \beta_2} \delta(t_1-t_2),
\end{eqnarray}
where $\alpha_1, \alpha_2 \in \{x,y,z\}$ and  $\beta_1, \beta_2 \in \{{\rm 0, r}\}$.
The Langevin equations are integrated by the leapfrog algorithm \cite{alle87,nogu11}.
In this study, we use $m= \zeta_{\rm 0}\tau_0$, $I=\zeta_{\rm r}\tau_0$, $\zeta_{\rm r}=\zeta_{\rm 0}\sigma^2$,
and $\Delta t=0.005\tau_0$ where $\tau_0= \zeta_0\sigma^2/k_{\rm B}T$.
The simulation results are displayed with a time unit of $\tau= r_{\rm {rod}}^2/D$,
where $D$ is the diffusion coefficient of the membrane particles in the tensionless membranes;
$D$ is calculated from the mean square displacement of the particles: $D\tau_0/r_{\rm {rod}}^2=0.001\pm0.0001$, 
so that $\tau=1000\tau_0$.
This time unit is estimated as $\tau \sim 10^{-4}s$
from $r_{\rm {rod}} \simeq 20$ nm and 
$D\simeq 4 \mu$m$^2$/s of transmembrane proteins \cite{rama09a}.

The assemblies of the isotropic inclusions and protein rods on
membrane tubes and vesicles with $N=2,400$ were investigated
at the inclusion density  $\phi_{\rm {iso}}=N_{\rm {iso}}/N=0.167$ and 
the rod density $\phi_{\rm {rod}}=N_{\rm {rod}}N_{\rm {sg}}/N=0.167$,
where $N_{\rm {iso}}$ and $N_{\rm {rod}}$ are the numbers of the isotropic inclusions and rods, respectively.
The replica exchange molecular dynamics \cite{huku96,okam04} for  $C_{\rm {iso}}$ or $C_{\rm {rod}}$ \cite{nogu14,nogu15b} 
is used to obtain the thermal equilibrium states.
Membrane rupture was investigated for vesicles with $N=9,600$ for $\phi_{\rm {rod}}\geq 0.5$.
The error bars are estimated from four and $10$ independent runs
for the replica exchange simulations and vesicle rupture, respectively.

The tube length is fixed in the longitudinal ($z$) direction and
periodic boundary conditions are employed.
The radius of the tube is $R_{\rm {cyl}}= 0.989 r_{\rm {rod}}$ for $L_z=4.8r_{\rm {rod}}$.
This tube radius is used, if not otherwise specified. 
In Sec.~\ref{sec:rodtube}, $L_z$ is also varied in order to investigate the tube radius dependence.
The radii of the vesicles
are $R_{\rm {ves}}= 1.54 r_{\rm {rod}}$ and $3.07 r_{\rm {rod}}$ at $N=2,400$ and $9,600$, respectively, in the absence of the rods.

\begin{figure}
\includegraphics{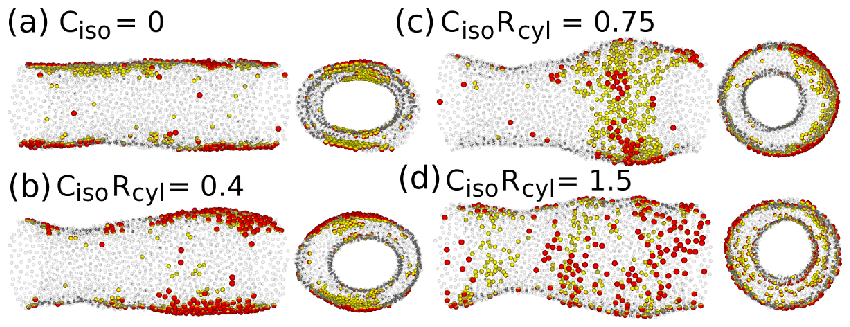}
\includegraphics{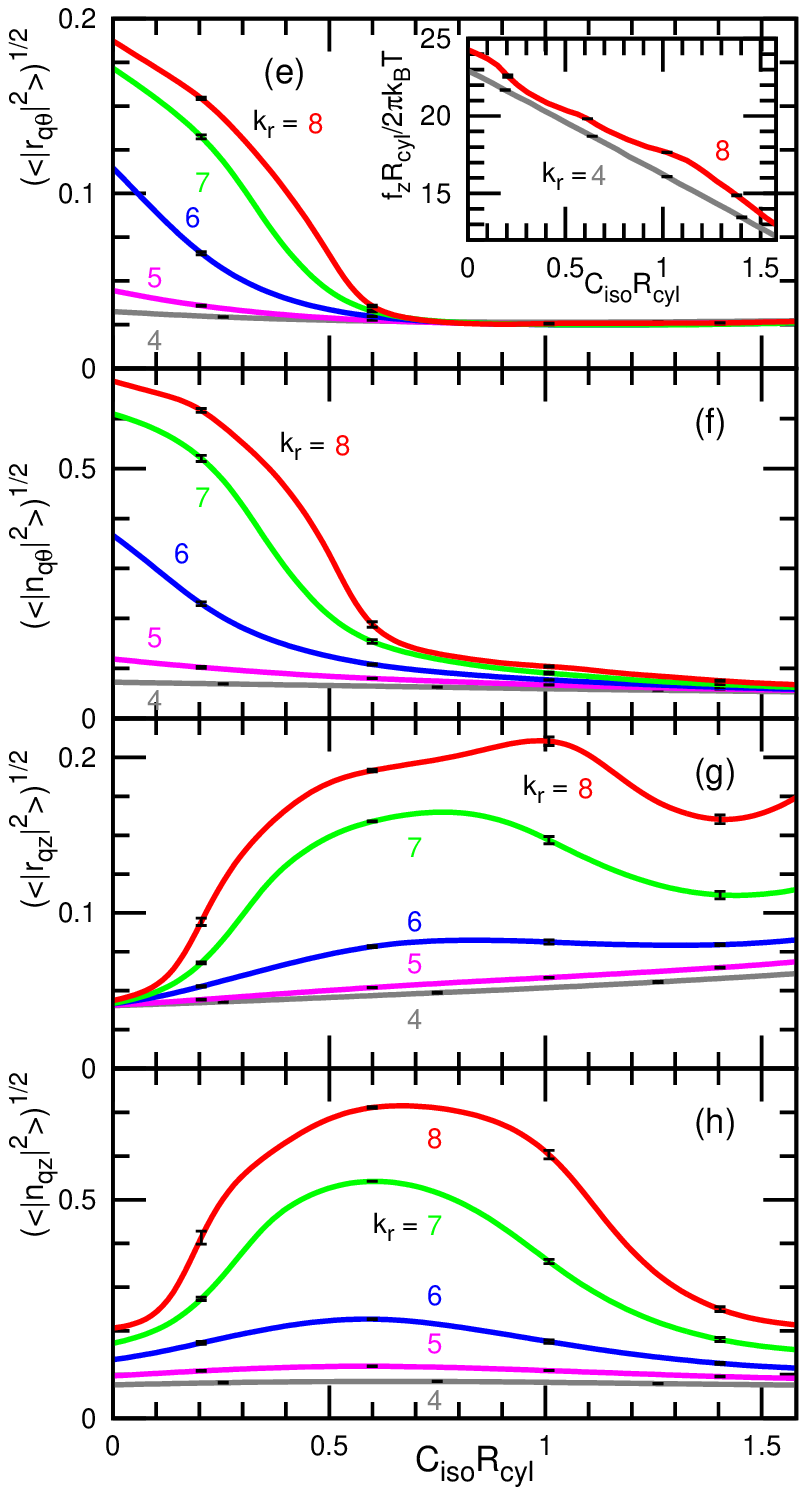}
\caption{
(Color online)
Deformation of membrane tubes induced by isotropic inclusions at $\phi_{\rm {iso}}=0.167$ and $N=2,400$.
(a)--(d) Snapshots for (a) $C_{\rm {iso}}R_{\rm {cyl}}=0$, (b) $0.4$, (c) $1$, and (d) $1.5$ at $k_{\rm r}=8$.
The front and side views are shown.
The inclusion  is displayed as a sphere whose halves are colored
in red (dark gray) and in yellow (light gray).
The orientation vector ${\bf u}_i$ lies along the direction from the 
yellow (light gray) to red (dark gray) hemispheres.
Transparent gray particles represent membrane particles.
(e)--(h) Fourier amplitudes of (e), (f) membrane shape and (g), (h) the inclusion densities
as functions of the spontaneous curvature $C_{\rm {iso}}$.
The amplitudes of the lowest Fourier mode along the azimuthal ($\theta$) and longitudinal ($z$) directions
are calculated for the membrane shape ($r_{q\theta}$ and $r_{qz}$) and densities ($n_{q\theta}$ and $n_{qz}$). 
The mean axial force $f_z$ of the membrane tube is shown in the inset of (e).
Error bars are displayed at several data points.
}
\label{fig:cyli}
\end{figure}

\begin{figure}
\includegraphics{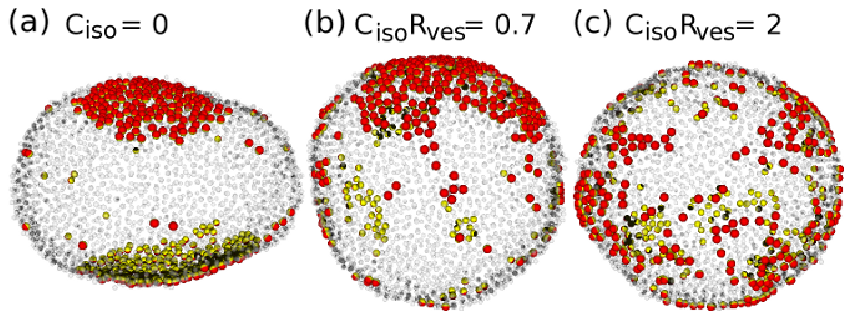}
\includegraphics{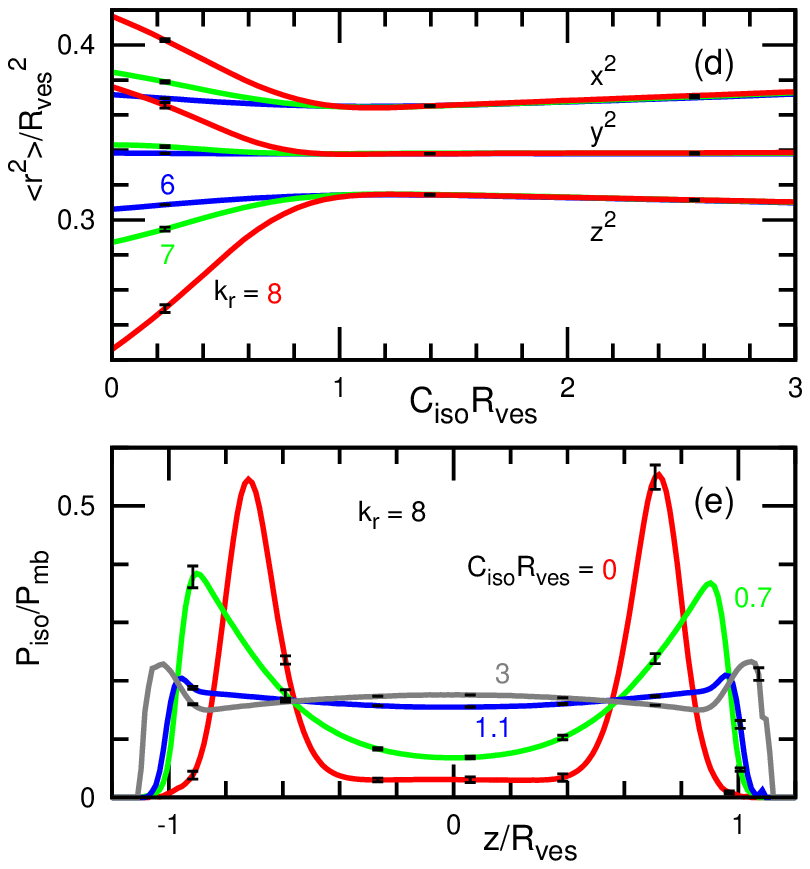}
\caption{
(Color online)
Vesicle deformation induced by isotropic inclusions at $\phi_{\rm {iso}}=0.167$ and $N=2,400$.
(a)--(c) Snapshots for (a) $C_{\rm {iso}}R_{\rm {ves}}=0$, (b) $0.7$, and (c) $2$ at $k_{\rm r}=8$.
(d) Three mean eigenvalues of the gyration tensor of the vesicle at $k_{\rm r}=6$, $7$, and $8$.
(e) Probability distribution of the inclusions along the $z$ axis
 for $C_{\rm {iso}}R_{\rm {ves}}=0$, $0.7$, $1.1$, and $3$ at $k_{\rm r}=8$.
Error bars are displayed at several data points.
}
\label{fig:sphi}
\end{figure}

\section{Isotropic inclusions}
\label{sec:iso}

First, we investigated the assembly of the isotropic inclusions 
in order to compare it with the rod assembly.
When the bending rigidity difference $k_{\rm r}$ between the inclusions and membrane is small,
the inclusions are isotropically distributed on a membrane tube and vesicle.
However, the inclusion assembly occurs at $k_{\rm r} \gtrsim 6$  (see Figs.~\ref{fig:cyli} and \ref{fig:sphi}).
The assembly regions have curvatures closer to the preferred curvature of the inclusions.
Hence, for hard inclusions, assembly reduces the bending energy, which is greater than the loss of the mixing entropy.

\subsection{Membrane Tube}
\label{sec:isotube}

For $C_{\rm {iso}}=0$ at $k_{\rm r} \gtrsim 6$,
the membrane tube deforms into an elliptic cylinder and
the inclusions assemble into two longitudinal domains in two flatter regions [see Fig.~\ref{fig:cyli}(a)].
The tips of the ellipse have a large curvature, 
but the inclusions in the flatter regions can have a low bending energy;
hence, the total bending energy is reduced.
This phase separation in the azimuthal direction is captured by the amplitudes of the Fourier modes [see Figs.~\ref{fig:cyli}(e)--(h)].
The lowest Fourier modes of the membrane shape and inclusion density along the azimuthal ($\theta$) direction
are given by  $r_{q\theta}= (1/N)\sum_i r_i \exp(-2\theta_i {\rm i})$
and $n_{q\theta}= (1/N_{\rm {rod}})\sum_i \exp(-2 \theta_i {\rm i})$, respectively,
where $\theta_i=\tan^{-1}(x_i/y_i)$.
In the axial ($z$) direction,
$r_{qz}= (1/N)\sum_i r_i \exp(-2\pi z_i {\rm i}/L_z)$
and  $n_{qz}= (1/N_{\rm {rod}})\sum_i \exp(-2\pi z_i {\rm i}/L_z)$.
With increasing $k_{\rm r}$, the amplitudes of $r_{q\theta}$ and $n_{q\theta}$ 
along the $\theta$ direction increase concurrently at  $C_{\rm {iso}}=0$.

As $C_{\rm {iso}}$ increases,
the membrane also deforms sigmoidally along the axial ($z$) direction 
and the inclusions become concentrated in the convex region [see Fig.~\ref{fig:cyli}(b)].
As a result, the Fourier amplitudes of $\theta$ and $z$ decrease and increase, respectively.
At $C_{\rm {iso}}R_{\rm {cyl}} \simeq 0.6$, the Fourier amplitudes of $\theta$ 
reach the values of the homogeneous membranes; the membrane deforms into a circular sigmoidal shape
and the inclusions are concentrated along a convex ring [see Fig.~\ref{fig:cyli}(c)].
With a further increase in $C_{\rm {iso}}$,
the inclusions become uniformly distributed and the membrane is less undulated in the axial direction
 [see Figs.~\ref{fig:cyli}(d), (g), and (h)].
The spontaneous curvatures between the inclusion pair and between the membrane particle and inclusion
are $C_{\rm {iso}}$ and $C_{\rm {iso}}/2$,
so that for $1 \lesssim C_{\rm {iso}}R_{\rm {cyl}}\lesssim 2$,
local membrane spontaneous curvature can match the curvature $1/R_{\rm {cyl}}$ of the membrane tube.

An even further increase in $C_{\rm {iso}}$ induces large membrane fluctuations, which induce
the contact of the membranes at an hourglass-like neck of the membrane tube.
This contact results in rupture of the membrane and formation of a spherical vesicle.
At $k_{\rm r}=8$, the rupture occurs at $C_{\rm {iso}}R_{\rm {cyl}}\gtrsim 2$.
With decreasing $k_{\rm r}$, larger values of $C_{\rm {iso}}$ are required for the rupture.

In a cylindrical tube of a homogeneous membrane, the bending energy yields an axial force
\begin{equation}
f_z = \frac{\partial F}{\partial L_z}\bigg|_A = 2\pi\kappa \Big(\frac{1}{R_{\rm cyl}} - C_0\Big), \label{eq:fz1}
\end{equation}
since an increase in the axial length results in a decrease in the cylindrical radius, 
i.e., an increase in the membrane mean curvature~\cite{shib11}.
At $k_{\rm r}=4$,
the force $f_z$ linearly decreases with increasing $C_{\rm {iso}}$ as shown in the inset of Fig.~\ref{fig:cyli}(e).
This indicates that the inclusions are homogeneously mixed in the membrane.
A similar linear dependence has previously been obtained for the density of anchored ideal-polymer chains~\cite{wu13}.
In contrast, for $k_{\rm r}=8$, the $f_z$--$C_{\rm {iso}}$ curve deviates from a straight line
due to the inclusion assembly. 
A larger shape change induces
greater deviation [compare the inset of Fig.~\ref{fig:cyli}(e) with Figs.~\ref{fig:cyli}(g) and (h)].

\subsection{Vesicle}
\label{sec:isoves}

Similarly to the membrane tube,
at $C_{\rm {iso}}=0$ and $k_{\rm r}=8$,
the isotropic inclusions deform a vesicle into an oblate shape 
and the inclusions are concentrated in the two flatter regions [see Fig.~\ref{fig:sphi}(a)].
As $C_{\rm {iso}}$ increases, the vesicle becomes more spherical and the inclusions are distributed more uniformly 
[see Figs.~\ref{fig:sphi}(b) and (c)].
For $C_{\rm {iso}}R_{\rm {ves}}\gtrsim 7$ at $k_{\rm r}=8$,
vesicle division occurs and two spherical vesicles are formed.

These changes in the vesicle shape and inclusion assembly can be captured by
the changes of the three principal lengths and inclusion distribution 
as shown in Figs.~\ref{fig:sphi}(d) and (e), respectively. 
The squared principal lengths $x^2, y^2$, and $z^2$ are the eigenvalues of the gyration tensor, 
$a_{\alpha\beta}= (1/N)\sum_j (\alpha_{j}-\alpha_{\rm G}) (\beta_{j}-\beta_{\rm G})$,
where $\alpha, \beta \in x,y,z$ and $\alpha_{\rm G}$ is the center of mass.
At $k_{\rm r}=6$, 
they are almost independent of $C_{\rm {iso}}$ 
so that the vesicle maintains its spherical shape.
Note that the differences among three eigenvalues at $k_{\rm r}=6$ are due to the thermal fluctuations.
At $k_{\rm r}=7$, small deviations are  recognized at $C_{\rm {iso}}\simeq 0$
and at $k_{\rm r}=8$, a clear decrease in $\langle z^2\rangle$ is obtained.
The inclusion distribution along the eigenvector of the smallest eigenvalue ($z$) is calculated
as $P_{\rm {iso}}/P_{\rm {mb}}$, where
$P_{\rm {iso}}$ and $P_{\rm {mb}}$ are probabilities of finding the inclusions and all particles at each $z$ bin, 
respectively.
The peaks of $P_{\rm {iso}}/P_{\rm {mb}}$ at both ends indicate the inclusion assembly on 
the flatter regions of 
the oblate vesicle 
[see the (red) solid line in Fig.~\ref{fig:sphi}(e)].
For $C_{\rm {iso}}R_{\rm {ves}}\gtrsim 1$, the inclusions are uniformly distributed.

\begin{figure}
\includegraphics{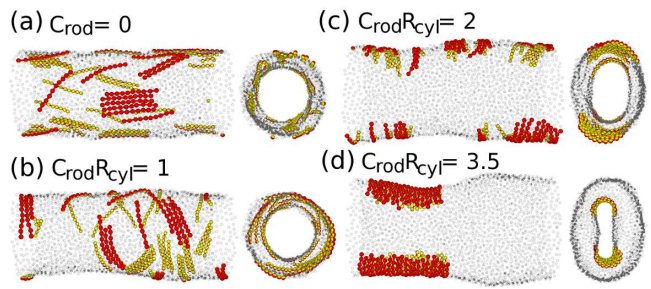}
\caption{
(Color online)
Snapshots of membrane tubes 
for (a) $C_{\rm {rod}}R_{\rm {cyl}}=0$, (b) $1$, (c) $2$, and (d) $3.5$ 
at $k_{\rm r}=12$ and $\phi_{\rm {rod}}=0.167$.
The front and side views are shown.
The protein rod is displayed as a chain of spheres whose halves are colored
in red (dark gray) and in yellow (light gray).
}
\label{fig:rcylsnap}
\end{figure}

\begin{figure}
\includegraphics{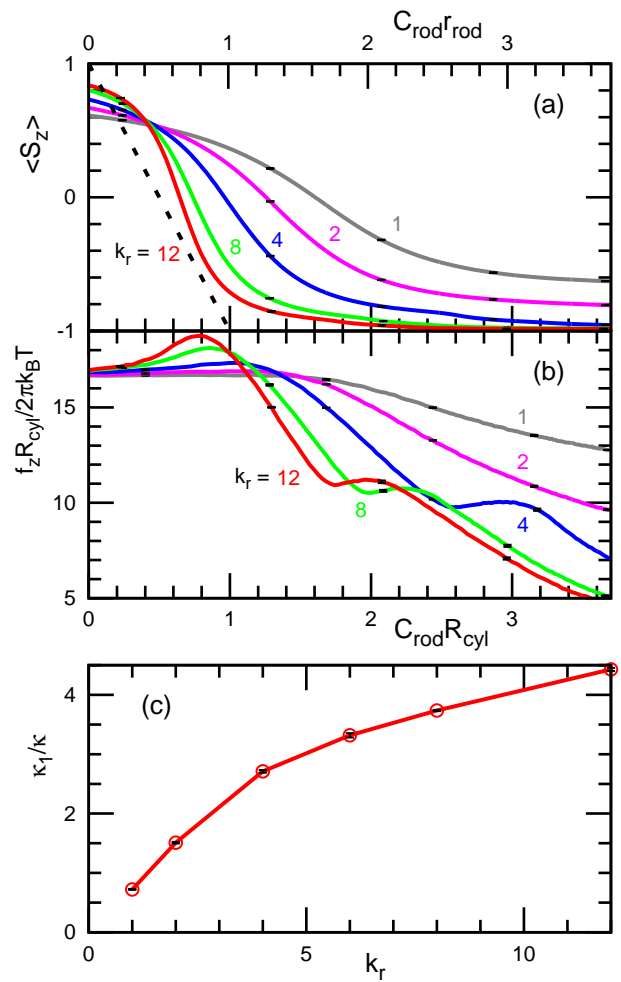}
\caption{
(Color online)
Rod orientation and axial force of the membrane tube
at $\phi_{\rm {rod}}=0.167$.
(a), (b) Rod curvature, $C_{\rm {rod}}$, dependence of  
(a) the orientation degree $S_z$ and (b) axial force $f_z$
at  $k_{\rm r}=1$, $2$, $4$, $8$, and $12$.
The (black) dashed line in (a) shows the relation of $S_z=1-2C_{\rm {rod}}R_{\rm {cyl}}$ 
for undeformable rods without thermal fluctuations.
(c) Effective bending rigidity of the rods estimated by Eq.~(\ref{eq:fz2}).
Error bars are displayed at several or all data points in (a) and (b) or (c), respectively.
}
\label{fig:rcylvir}
\end{figure}

\begin{figure}[t]
\includegraphics{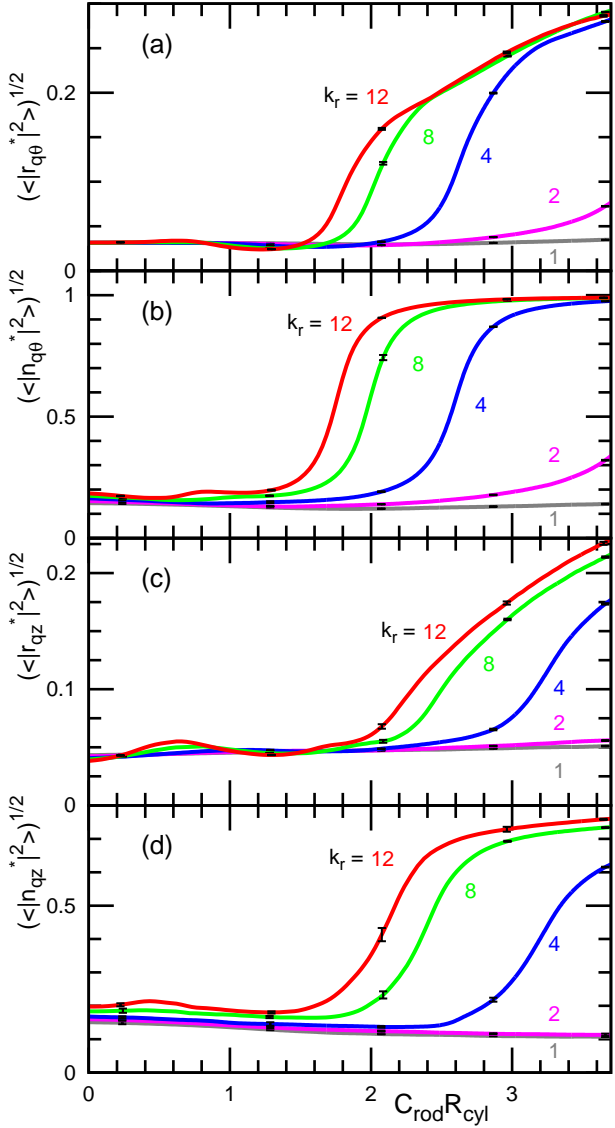}
\caption{
(Color online)
Rod curvature, $C_{\rm {rod}}$, dependence of  
Fourier amplitudes of (a), (c) membrane shape and (b), (d) the rod densities
at  $\phi_{\rm {rod}}=0.167$ and  $k_{\rm r}=1$, $2$, $4$, $8$, and $12$.
The Fourier amplitudes are normalized by the values at $C_{\rm {rod}}=0$ 
(denoted by the superscript $*$).
Error bars are displayed at several data points.
}
\label{fig:rcylrn}
\end{figure}

\begin{figure}[t]
\includegraphics{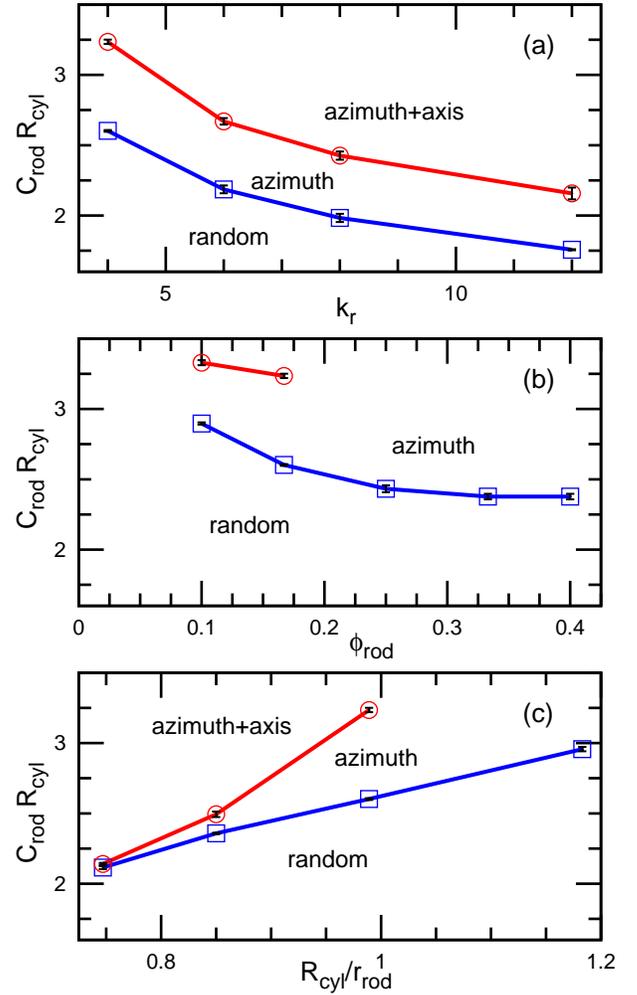}
\caption{
(Color online)
Phase diagrams of the membrane tube.
(a) $C_{\rm {rod}}$--$k_{\rm r}$ diagram at $\phi_{\rm {rod}}=0.167$ and $R_{\rm {cyl}}=0.989r_{\rm {rod}}$.
(b) $C_{\rm {rod}}$--$\phi_{\rm {rod}}$ diagram at $k_{\rm {r}}=4$ and $R_{\rm {cyl}}=0.989r_{\rm {rod}}$.
(c) $C_{\rm {rod}}$--$R_{\rm {cyl}}/r_{\rm {rod}}$ diagram at $k_{\rm {r}}=4$ and  $\phi_{\rm {rod}}=0.167$.
}
\label{fig:pdeq}
\end{figure}

Compared to the assembly of the anisotropic rods, as described in the next section,
the inclusions assemble weakly and the shapes of the membrane tubes and vesicles fluctuate greatly,
and for small values of $C_{\rm {iso}}$, a single domain, instead of two domains, is occasionally formed.
High bending rigidity of the inclusion is required for assembly in the absence of 
direct attraction between the inclusions.

\section{Membrane tube with protein rods}
\label{sec:rodtube}

The protein rods exhibit a two-step assembly with increasing rod curvature $C_{\rm {rod}}$ 
(see Figs.~\ref{fig:rcylsnap}--\ref{fig:pdeq}).
In our previous papers~\cite{nogu14,nogu15b}, we reported the assembly at $k_{\rm r}=4$.
Here, we show the $k_{\rm r}$ dependence and summarize the assembly processes.

At $C_{\rm {rod}}=0$, the protein rods are oriented along the axial ($z$) direction
and uniformly distributed  in the membrane tube [see Fig.~\ref{fig:rcylsnap}(a)].
As $C_{\rm {rod}}$ increases, the rod orientation changes to the azimuthal ($\theta$) direction
and the mean orientational order parameter $\langle S_z\rangle$ decreases from $1$ to $-1$,
where the orientational order parameter is defined as $S_z = (1/N_{\rm rod})\sum_i ( 2{s_{i,z}}^2-1 )$
 [see Fig.~\ref{fig:rcylvir}(a)].
At larger $k_{\rm r}$, $\langle S_z\rangle$  decreases more rapidly.
For the limit of $k_{\rm r}\to \infty$, i.e., undeformable rods, 
the curvature along the rod axis is exactly $C_{\rm {rod}}$.
If the thermal fluctuations are neglected, a linear relation $S_z=1-2C_{\rm {rod}}R_{\rm {cyl}}$ 
is obtained for $0\leq C_{\rm {rod}}R_{\rm {cyl}}\leq 1$.
As $k_{\rm r}$ increases, $\langle S_z\rangle$ approaches this linear relation, but
 relatively large deviations remain at $\langle S_z\rangle \simeq \pm 1$.
These deviations are likely due to membrane undulations, 
which allow orientation fluctuations even for undeformable rods.

The axial force $f_z$ behaves differently from the close-to-linear dependence of the isotropic inclusions [compare Fig.~\ref{fig:rcylvir}(b)
with the inset of Fig.~\ref{fig:cyli}(e)].
During orientation changes, $f_z$ is almost constant for small $k_{\rm r}$,
while it increases slightly for large $k_{\rm r}$.
This increase may be due to entropy reduction by tilted rods,
since the tilted rods suppress membrane undulation in both the $z$ and $\theta$ directions.
In this region ($0\leq C_{\rm {rod}}R_{\rm {cyl}}\lesssim 1$),
 changes in the Fourier amplitudes are very small in both directions for all values of $k_{\rm r}$ 
(see Fig.~\ref{fig:rcylrn}).
Therefore, the membrane shapes and axial stress are modified only a little by the rods in this region.

With a further increase in $C_{\rm {rod}}$, $f_z$ decreases linearly until the azimuthal assembly commences.
When the rods are assumed to be completely oriented in the azimuthal direction,
the axial force is given by~\cite{nogu15b},
\begin{equation}
f_z = \frac{2\pi\kappa}{R_{\rm cyl}} +  2\pi\phi_{\rm {rod}} \Big( \frac{\kappa_1-\kappa}{R_{\rm cyl}} - \kappa_1 C_{\rm {rod}}\Big) . \label{eq:fz2}
\end{equation}
The first term is the force in the absence of rods.
The second term is the force generated by the rods,
which is proportional to $\phi_{\rm {rod}}$ and linear with respect to $C_{\rm {rod}}$.
The effective bending rigidity $\kappa_1$ of the rods is estimated from the slope of $f_z$--$C_{\rm {rod}}$ curves
in the linear-decrease regions
as shown in Fig.~\ref{fig:rcylvir}(c).
$\kappa_1/\kappa$ increases with $k_{\rm r}$ but is not linear with $k_{\rm r}$.
This is because the orientation is not completely in the azimuthal direction 
and the interactions between rods and neighboring membrane particles are also involved in the  bending deformation along the rod axis.

With an even further increase in $C_{\rm {rod}}$, 
the rods assemble along the azimuthal direction
and the membrane deforms into an elliptic tube [see Fig.~\ref{fig:rcylsnap}(c)].
As $C_{\rm {rod}}$ increases more, the rod assembly also occurs along the axial direction
[see Fig.~\ref{fig:rcylsnap}(d)].
The increases in the Fourier amplitudes of the azimuthal ($\theta$) and axial ($z$) modes 
indicate the  azimuthal and axial assemblies, respectively
(see Fig.~\ref{fig:rcylrn}).
With increasing $k_{\rm r}$,
both assemblies occur at smaller $C_{\rm {rod}}$ [see Figs.~\ref{fig:rcylrn} and \ref{fig:pdeq}(a)].
The curvatures $C_{\rm {rod}}$ of the azimuthal and axial assembly points are determined 
by the inflection points of $(\langle |n_{q\theta}|^2 \rangle)^{1/2}$ 
and $(\langle |n_{qz}|^2 \rangle)^{1/2}$, respectively.
The assembly is enhanced by the large rod stiffness in a manner similar to that of the isotropic inclusions.
Thus, rod elasticity is one of the important factors that determine the assembly curvatures.

The phase diagram for the rod density $\phi_{\rm {rod}}$ and the tube radius $R_{\rm {cyl}}$
are shown in Figs.~\ref{fig:pdeq}(b) and (c), respectively.
For the azimuthal assembly, 
$\phi_{\rm {rod}}$ gives an effect that is very similar to that of the rod stiffness $k_{\rm r}$.
However, the axial assembly is different.
At a large density ($\phi_{\rm {rod}}\gtrsim 0.25$), 
axial assembly does not occur, since the elliptic edges are filled by the rods \cite{nogu15b}.
As $\phi_{\rm {rod}}$ increases further,
the membrane deforms into a triangular or other polygonal tube, instead of the elliptic tube, and
the rods assemble at the edges of the polygonal tube \cite{nogu15b}.
As $R_{\rm {cyl}}$ increases, a slightly larger $C_{\rm {rod}}R_{\rm {cyl}}$ is needed 
for assembly in both directions.
At a large tube radius ($R_{\rm {cyl}}/r_{\rm {rod}} \gtrsim 1$), 
 axial assembly does not occur for the same reason as for a large $\phi_{\rm {rod}}$.
The length $L_z$ decreases as $L_z \propto 1/R_{\rm {cyl}}$ for the constant membrane area,
so that the elliptic edges can be filled by a smaller number of rods.

\begin{figure}
\includegraphics[width=8cm]{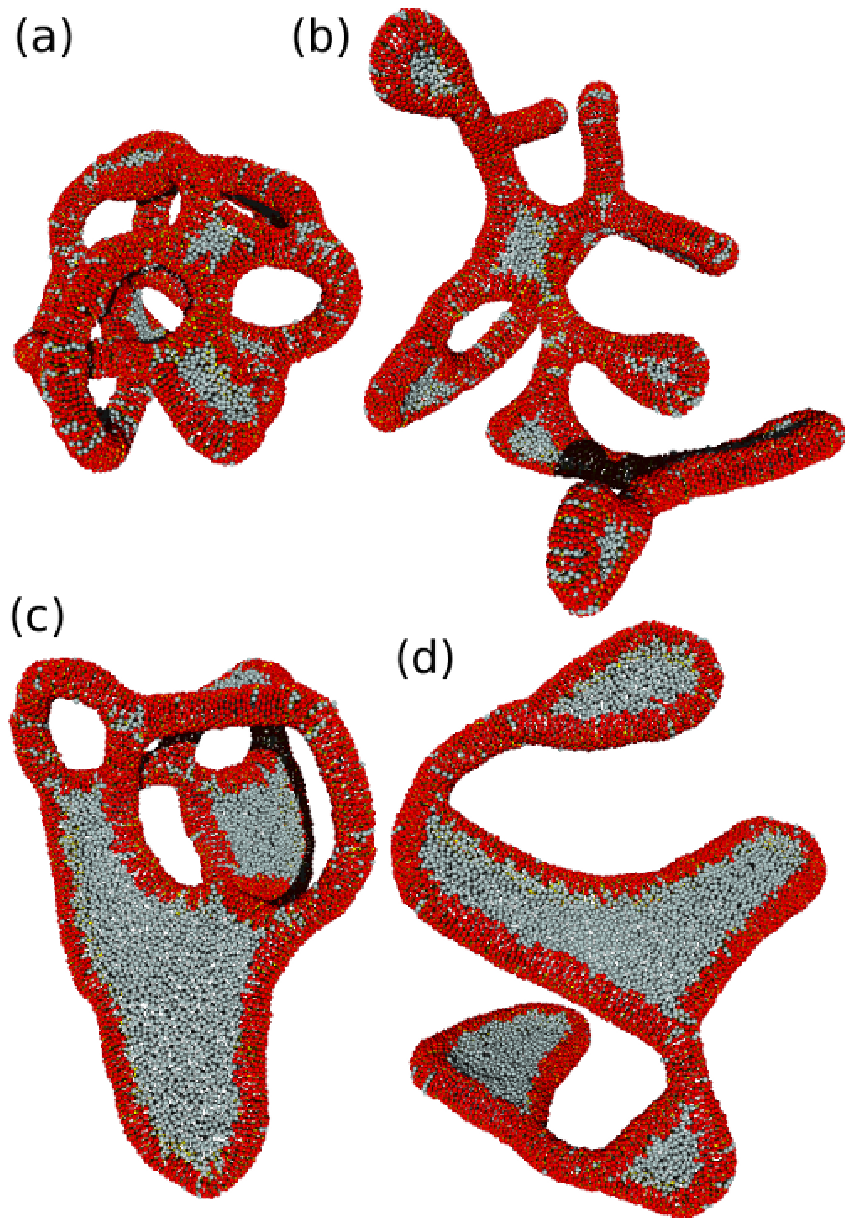}
\includegraphics{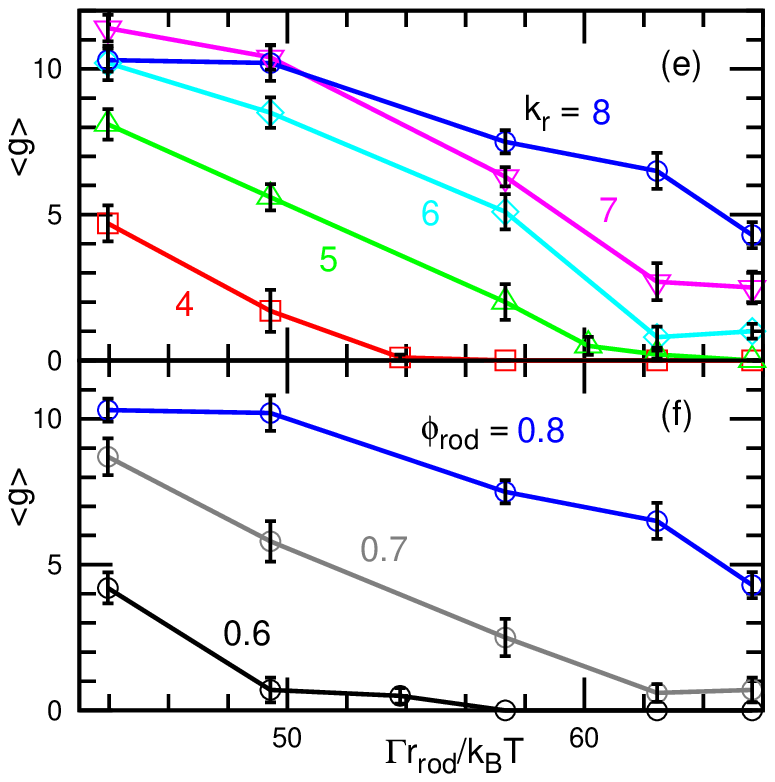}
\caption{
(Color online)
Membrane rupture at $C_{\rm {rod}}R_{\rm {ves}}=6.2$ and $N=9,600$.
(a)--(d) Snapshots of vesicles.
(a)  Genus-11 vesicle at $\Gamma r_{\rm {rod}}/k_{\rm B}T=49$, $k_{\rm r}=6$, and $\phi_{\rm {rod}}=0.8$.
(b)  Genus-1 vesicle at $\Gamma r_{\rm {rod}}/k_{\rm B}T=66$, $k_{\rm r}=6$, and $\phi_{\rm {rod}}=0.8$.
(c)  Genus-5 vesicle at $\Gamma r_{\rm {rod}}/k_{\rm B}T=44$, $k_{\rm r}=8$, and $\phi_{\rm {rod}}=0.6$.
(d)  Genus-0 vesicle at $\Gamma r_{\rm {rod}}/k_{\rm B}T=57$, $k_{\rm r}=8$, and $\phi_{\rm {rod}}=0.6$.
Membrane particles are displayed by nontransparent gray particles.
(e)--(f) Mean number of the genus $\langle g \rangle$
of vesicles as functions of the line tension $\Gamma$.
(e) $k_{\rm r}=4$, $5$, $6$, $7$, and $8$ for $\phi_{\rm {rod}}=0.8$.
(f) $\phi_{\rm {rod}}=0.6$, $0.7$, and $0.8$ for $k_{\rm r}=8$.
}
\label{fig:rup_out}
\end{figure}

\begin{figure}[!ht]
\includegraphics[width=8cm]{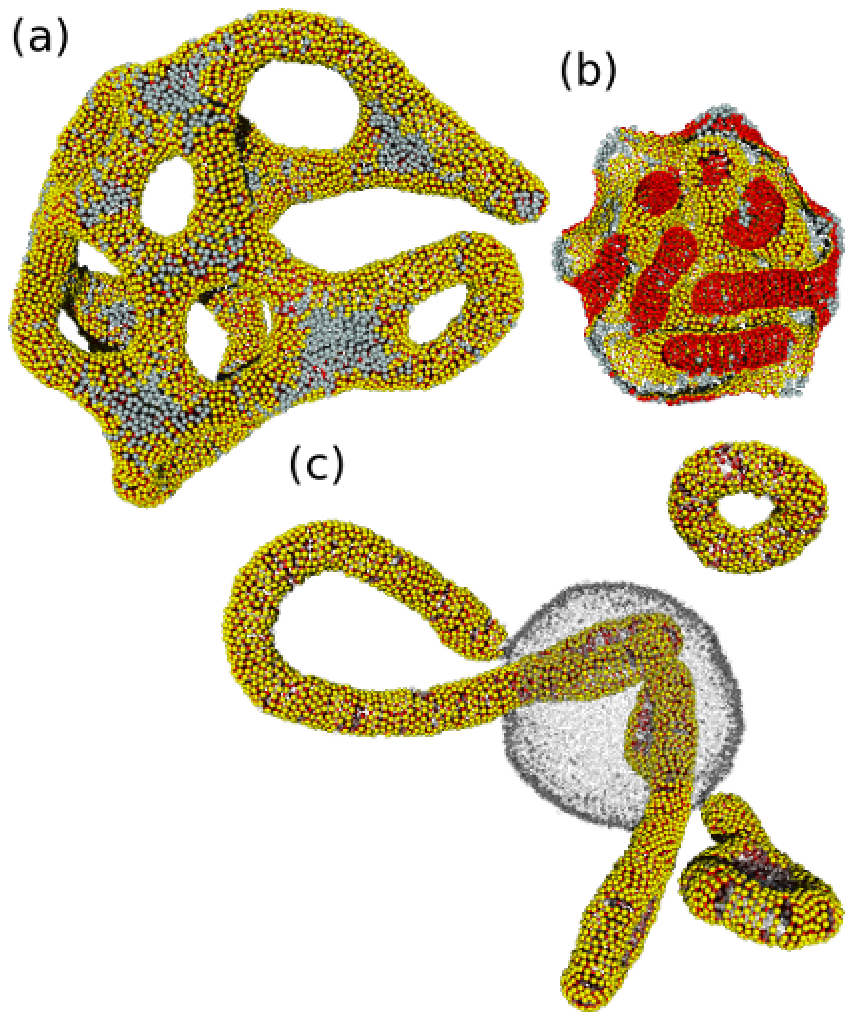}
\includegraphics{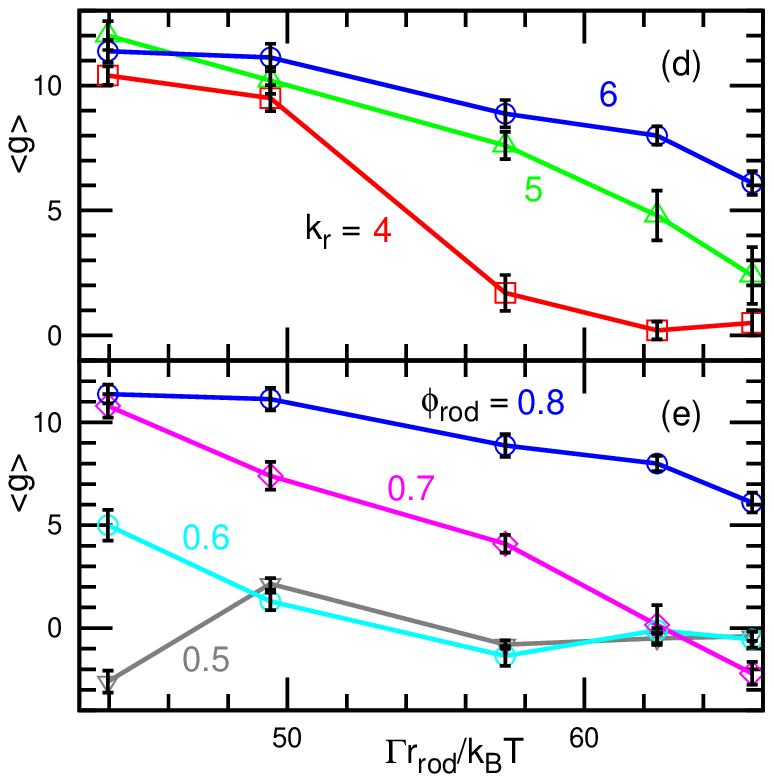}
\caption{
(Color online)
Membrane rupture at $C_{\rm {rod}}R_{\rm {ves}}=-6.2$ and $N=9,600$.
(a)--(c) Snapshots of vesicles.
(a)  Genus-11 vesicle at $\Gamma r_{\rm {rod}}/k_{\rm B}T=49$, $k_{\rm r}=4$, and $\phi_{\rm {rod}}=0.8$.
(b)  Invaginated genus-0 vesicle at $\Gamma r_{\rm {rod}}/k_{\rm B}T=62$, $k_{\rm r}=4$, and $\phi_{\rm {rod}}=0.8$.
(c)  Ruptured membrane at $\Gamma r_{\rm {rod}}/k_{\rm B}T=49$, $k_{\rm r}=6$, and $\phi_{\rm {rod}}=0.5$.
Membrane particles are displayed by nontransparent or transparent gray particles in (a) and (b) or (c), respectively,
for clarity.
(d)--(e) Mean number of the genus $\langle g \rangle$
of vesicles as functions of the line tension $\Gamma$.
(d) $k_{\rm r}=4$, $5$, and $6$ for $\phi_{\rm {rod}}=0.8$.
(e) $\phi_{\rm {rod}}=0.5$, $0.6$, $0.7$, and $0.8$ for $k_{\rm r}=6$.
}
\label{fig:rup_in}
\end{figure}

\begin{figure}
\includegraphics[width=8cm]{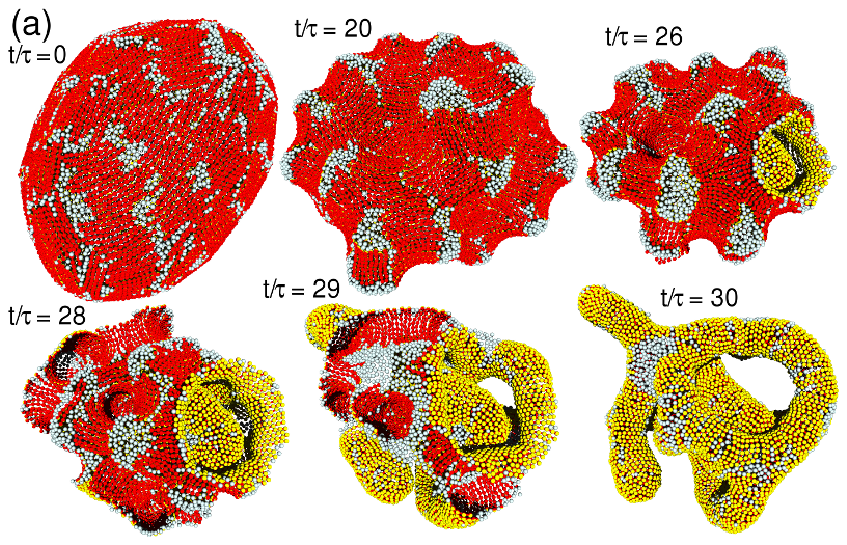}
\includegraphics{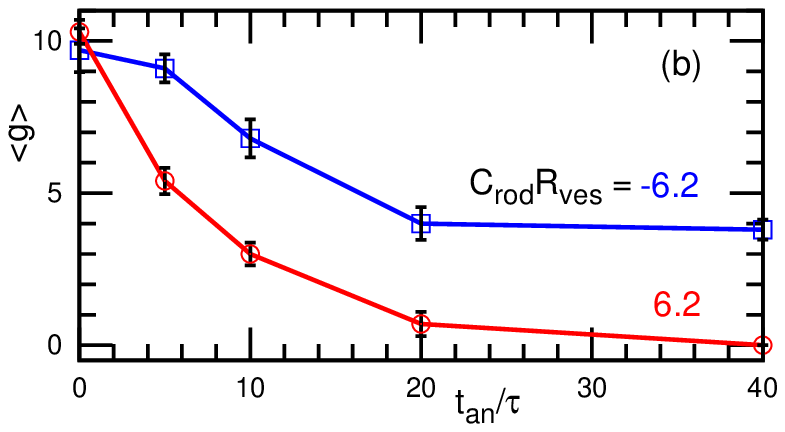}
\caption{
(Color online)
Annealing time, $t_{\rm {an}}$, dependence of
membrane rupture at   $\Gamma r_{\rm {rod}}/k_{\rm B}T=44$, $k_{\rm r}=8$, $\phi_{\rm {rod}}=0.8$, and $N=9,600$.
(a) Sequential snapshots of a vesicle for $t/\tau=0$, $20$, $26$, $28$, $29$, and $30$
at $C_{\rm {rod}}R_{\rm {ves}}=-6.2$ and $t_{\rm {an}}/\tau=40$.
Membrane particles are displayed by nontransparent gray particles.
(b) Mean number of the genus $\langle g \rangle$ as functions of $t_{\rm {an}}$
for $C_{\rm {rod}}R_{\rm {ves}}=-6.2$ and $6.2$.
}
\label{fig:rup_an}
\end{figure}

Protein rods with large rod curvatures form a tight assembly 
in the membrane tubes as well as in the vesicles~\cite{nogu14,nogu15b}. 
It differs from the assembly of isotropic inclusions into flat membranes, 
which requires a larger bending elasticity.
The rod assembly forms a saddle shape in the membrane tube.
In contrast, such a saddle membrane is not stabilized by isotropic inclusions 
with a large positive spontaneous curvature
since its mean curvature is small.

\section{Vesicle rupture}
\label{sec:rup}

Membranes can rupture under a large stress.
It is determined by the competition between the membrane deforming force (by the protein rods in this study)
and the line tension $\Gamma$ of the membrane edge.
For example, 
under a large positive surface tension $\gamma$,
a membrane pore can expand until the pore radius reaches the stable value
 $R_{\rm {pore}}=\gamma/\Gamma$ \cite{tolp04,nogu06}.
A vesicle can spontaneously transform into a disk-shaped flat membrane when $\Gamma < (2\kappa+\bar\kappa)/R_{\rm {ves}}$ \cite{from83,hu12,naka15}.
Here, we use sufficiently large $\Gamma$ to prevent membrane rupture in the absence of rods.

Figures~\ref{fig:rup_out}--\ref{fig:rup_an} show vesicle shapes resulting from membrane rupture by rod bending.
Rupture changes vesicle topology.
Figure~\ref{fig:rup_out} shows the vesicle shapes as the rod curvature changed suddenly from $C_{\rm {rod}}R_{\rm {ves}}=0$ to $6.2$.
As the edge tension $\Gamma$ decreases, the membrane is ruptured more frequently
and vesicles with higher genus $g$ are formed [see Figs.~\ref{fig:rup_out}(e) and (f)].
As the rod stiffness $k_{\rm r}$ or rod density $\phi_{\rm {rod}}$ increases,
the rods deform the membrane more rapidly and
 membrane rupture is enhanced.
For $\langle g \rangle \lesssim 8$, $\langle g \rangle$ depends linearly on $\Gamma$
while it is saturated at  $\langle g \rangle \simeq 10$.
A greater number of $\langle g \rangle$ indicates the occurrence of more rupture.
The obtained values of $g$ have a narrow distribution as indicated in small error bars that represent standard errors for $10$ samples.
The saturation of  $\langle g \rangle$  
is likely caused by the upper limit of genus, $g \simeq 13$, which is determined by the vesicle size.
A higher genus can be obtained for larger vesicles.
The obtained high-genus vesicles, shown in Fig.~\ref{fig:rup_out}(a),
agree with the previously reported shapes in simulations and experiments \cite{ayto09,simu13a}. 
Here, we have clarified that the edge tension and rod stiffness and density
are important factors for vesicle rupture.
At low $\phi_{\rm {rod}}$, 
rod-free membranes are phase-separated and form flat regions
while the rod assemblies form tubes and semicylindrical edges of the flat membranes [see Figs.~\ref{fig:rup_out}(c) and (d)].
A similar coexistence of tubular networks and flat membranes is seen in endoplasmic reticulum (ER)~\cite{shib09}.
When $C_{\rm {rod}}$ changes slowly rather than rapidly,
 vesicle rupture is suppressed.
The lower (red) line in Fig.~\ref{fig:rup_an}(b) shows $\langle g \rangle$ as a function of the annealing time
$t_{\rm {an}}$, for which $C_{\rm {rod}}(t)$ is changed linearly from $C_{\rm {rod}}R_{\rm {ves}}=0$ to $6.2$.
No vesicles are ruptured at $t_{\rm {an}}=40\tau$ ($\sim 4$ ms). 
Thus, the rapid adsorption of proteins onto the membrane is also important for obtaining vesicle rupture in experiments.

We also investigated vesicle rupture by protein rods 
with a curvature opposite to that of the vesicle curvature (see Figs.~\ref{fig:rup_in} and \ref{fig:rup_an}).
At first, the rods induce many tubular invaginations into the inside of the vesicle [see Fig.~\ref{fig:rup_in}(b)].
For large $\Gamma$ with relatively small $k_{\rm r}$, the membrane is not ruptured 
and densely-packed invaginations remain.
Similar invaginations were observed by electron microscopy for liposomes with I-BAR proteins~\cite{matt07}.
For small $\Gamma$ or large $k_{\rm r}$, 
the membrane is ruptured and subsequently the inside surface of the vesicle turns to the outside [see Fig.~\ref{fig:rup_an}(a)].
At high $\phi_{\rm {rod}}$, this inversion (inside out) occurs completely and a high-genus vesicle is formed [see Fig.~\ref{fig:rup_in}(a)]. 
However, at low  $\phi_{\rm {rod}}$, the inversion is only partial and 
rod-free membrane regions are not inverted, leading to division and partial connection of the membranes. 
In the membrane shown in Fig.~\ref{fig:rup_in}(c), the rod-free membrane forms a pored vesicle and
inverted membrane tubes partially remain in the vesicle. The ends of the tubes can be connected to the vesicle,
but some of them are eventually pinched off. For genus estimation, complete membrane fission and 
partial connection are counted as $-1$ and $-0.5$, respectively
[$g=-2$ in the case of Fig.~\ref{fig:rup_in}(c)]. 
Since the bending energy of an initially spherical vesicle is larger than the positive rod curvature $C_{\rm {rod}}$
with the same amplitude, membrane rupture occurs at smaller  $\phi_{\rm {rod}}$, $k_{\rm r}$, or larger  $\Gamma$.
Even when $C_{\rm {rod}}(t)$ is annealed slowly, high-genus vesicles are still formed [see Fig.~\ref{fig:rup_an}(b)].

\section{Summary}
\label{sec:sum}

We have investigated the shape transformation of vesicles and membrane tubes induced by protein rod assembly.
As the rod curvature $C_{\rm {rod}}$ increases, the protein rods in the membrane tube assemble by two steps;
first in the azimuthal direction and next in the longitudinal direction.
These assemblies occur at lower $C_{\rm {rod}}$ for stiffer rods and/or higher rod density.
Compared to the anisotropic rods,  laterally isotropic inclusions assemble weakly.
The isotropic inclusions assemble only when they induce very large bending rigidity locally; 
The inclusions with small spontaneous curvatures assemble in flatter regions of an elliptic membrane tube and oblate vesicle, 
while inclusions with large spontaneous curvatures induce membrane fission into vesicles.
The rod--rod excluded-volume interactions make
protein rods that are closer than $r_{\rm {rod}}/2$ align in the same orientation.
Thus, the elongated shapes of membrane-reshaping proteins also assist protein assembly.

When the protein rods induce a large membrane stress,
the membrane can be ruptured and high-genus vesicles form.
We have clarified that membrane rupture is induced by the large bending stiffness of the rods, high density, 
rapid protein adhesion, and/or low line tension of the membrane edge.
Thus, the choice of lipids is also important.
When $C_{\rm {rod}}$ is negative with respect to the initial vesicle curvature, 
the membrane inversion also results in vesicle division as well as in the formation of high-genus vesicles.
Our simulation results suggest that rapid exposure of liposomes to a protein solution is a key factor for observing 
high-genus liposomes due to protein-induced membrane rupture.
Vesicle inversion was previously observed during lysis of a liposome by detergents~\cite{nomu01}.
Here, we have demonstrated that the protein adhesion can also induce the vesicle inversion.

Here, we considered that the spontaneous (side) curvature and bending rigidity of the rods perpendicular to the rod axis 
are the same as the other membrane regions. Excluded-volume and other interactions between the proteins 
or between the protein and membrane can generate effective side spontaneous curvatures.
When the rod and side curvatures are in opposite directions, saddle-shaped membranes, such as
egg-carton \cite{domm99,domm02}, ring, and network structures \cite{nogu16}, can be stabilized.
Thus, anisotropic inclusions can induce much more variety in membrane structures than isotropic inclusions.

\begin{acknowledgments}
The replica exchange simulations were
carried out on SGI Altix ICE 8400EX and ICE XA
at ISSP Supercomputer Center, University of Tokyo. 
This work was partially supported by a Grant-in-Aid for Scientific Research on Innovative Areas 
``Fluctuation \& Structure'' (No. 25103010) from 
the Ministry of Education, Culture, Sports, Science, and Technology of Japan.
\end{acknowledgments}


\begin{thebibliography}{67}
\expandafter\ifx\csname natexlab\endcsname\relax\def\natexlab#1{#1}\fi
\expandafter\ifx\csname bibnamefont\endcsname\relax
  \def\bibnamefont#1{#1}\fi
\expandafter\ifx\csname bibfnamefont\endcsname\relax
  \def\bibfnamefont#1{#1}\fi
\expandafter\ifx\csname citenamefont\endcsname\relax
  \def\citenamefont#1{#1}\fi
\expandafter\ifx\csname url\endcsname\relax
  \def\url#1{\texttt{#1}}\fi
\expandafter\ifx\csname urlprefix\endcsname\relax\def\urlprefix{URL }\fi
\providecommand{\bibinfo}[2]{#2}
\providecommand{\eprint}[2][]{\url{#2}}

\bibitem[{\citenamefont{McMahon and Gallop}(2005)}]{mcma05}
\bibinfo{author}{\bibfnamefont{H.~T.} \bibnamefont{McMahon}} \bibnamefont{and}
  \bibinfo{author}{\bibfnamefont{J.~L.} \bibnamefont{Gallop}},
  \bibinfo{journal}{Nature} \textbf{\bibinfo{volume}{438}},
  \bibinfo{pages}{590} (\bibinfo{year}{2005}).

\bibitem[{\citenamefont{Shibata et~al.}(2009)\citenamefont{Shibata, Hu, Kozlov,
  and Rapoport}}]{shib09}
\bibinfo{author}{\bibfnamefont{Y.}~\bibnamefont{Shibata}},
  \bibinfo{author}{\bibfnamefont{J.}~\bibnamefont{Hu}},
  \bibinfo{author}{\bibfnamefont{M.~M.} \bibnamefont{Kozlov}},
  \bibnamefont{and} \bibinfo{author}{\bibfnamefont{T.~A.}
  \bibnamefont{Rapoport}}, \bibinfo{journal}{Annu.\ Rev.\ Cell\ Dev.\ Biol.}
  \textbf{\bibinfo{volume}{25}}, \bibinfo{pages}{329} (\bibinfo{year}{2009}).

\bibitem[{\citenamefont{Drin and Antonny}(2010)}]{drin10}
\bibinfo{author}{\bibfnamefont{G.}~\bibnamefont{Drin}} \bibnamefont{and}
  \bibinfo{author}{\bibfnamefont{B.}~\bibnamefont{Antonny}},
  \bibinfo{journal}{FEBS Lett.} \textbf{\bibinfo{volume}{584}},
  \bibinfo{pages}{1840} (\bibinfo{year}{2010}).

\bibitem[{\citenamefont{Baumgart et~al.}(2011)\citenamefont{Baumgart, Capraro,
  Zhu, and Das}}]{baum11}
\bibinfo{author}{\bibfnamefont{T.}~\bibnamefont{Baumgart}},
  \bibinfo{author}{\bibfnamefont{B.~R.} \bibnamefont{Capraro}},
  \bibinfo{author}{\bibfnamefont{C.}~\bibnamefont{Zhu}}, \bibnamefont{and}
  \bibinfo{author}{\bibfnamefont{S.~L.} \bibnamefont{Das}},
  \bibinfo{journal}{Annu. Rev. Phys. Chem.} \textbf{\bibinfo{volume}{62}},
  \bibinfo{pages}{483} (\bibinfo{year}{2011}).

\bibitem[{\citenamefont{Johannes et~al.}(2014)\citenamefont{Johannes, Wunder,
  and Bassereau}}]{joha14}
\bibinfo{author}{\bibfnamefont{L.}~\bibnamefont{Johannes}},
  \bibinfo{author}{\bibfnamefont{C.}~\bibnamefont{Wunder}}, \bibnamefont{and}
  \bibinfo{author}{\bibfnamefont{P.}~\bibnamefont{Bassereau}},
  \bibinfo{journal}{Cold Spring Harbor Perspect. Biol.}
  \textbf{\bibinfo{volume}{6}}, \bibinfo{pages}{a016741}
  (\bibinfo{year}{2014}).

\bibitem[{\citenamefont{McMahon and Boucrot}(2015)}]{mcma15}
\bibinfo{author}{\bibfnamefont{H.~T.} \bibnamefont{McMahon}} \bibnamefont{and}
  \bibinfo{author}{\bibfnamefont{E.}~\bibnamefont{Boucrot}},
  \bibinfo{journal}{J. Cell Sci.} \textbf{\bibinfo{volume}{128}},
  \bibinfo{pages}{1065} (\bibinfo{year}{2015}).

\bibitem[{\citenamefont{Itoh and {De Camilli}}(2006)}]{itoh06}
\bibinfo{author}{\bibfnamefont{T.}~\bibnamefont{Itoh}} \bibnamefont{and}
  \bibinfo{author}{\bibfnamefont{P.}~\bibnamefont{{De Camilli}}},
  \bibinfo{journal}{Biochim.\ Biophys.\ Acta} \textbf{\bibinfo{volume}{1761}},
  \bibinfo{pages}{897} (\bibinfo{year}{2006}).

\bibitem[{\citenamefont{Masuda and Mochizuki}(2010)}]{masu10}
\bibinfo{author}{\bibfnamefont{M.}~\bibnamefont{Masuda}} \bibnamefont{and}
  \bibinfo{author}{\bibfnamefont{N.}~\bibnamefont{Mochizuki}},
  \bibinfo{journal}{Semin. Cell Dev. Biol.} \textbf{\bibinfo{volume}{21}},
  \bibinfo{pages}{391} (\bibinfo{year}{2010}).

\bibitem[{\citenamefont{Zhao et~al.}(2011)\citenamefont{Zhao,
  Pyk{\"a}l{\"a}inen, and Lappalainen}}]{zhao11}
\bibinfo{author}{\bibfnamefont{H.}~\bibnamefont{Zhao}},
  \bibinfo{author}{\bibfnamefont{A.}~\bibnamefont{Pyk{\"a}l{\"a}inen}},
  \bibnamefont{and}
  \bibinfo{author}{\bibfnamefont{P.}~\bibnamefont{Lappalainen}},
  \bibinfo{journal}{Curr. Opin. Cell Biol.} \textbf{\bibinfo{volume}{23}},
  \bibinfo{pages}{14} (\bibinfo{year}{2011}).

\bibitem[{\citenamefont{Mim and Unger}(2012)}]{mim12a}
\bibinfo{author}{\bibfnamefont{C.}~\bibnamefont{Mim}} \bibnamefont{and}
  \bibinfo{author}{\bibfnamefont{V.~M.} \bibnamefont{Unger}},
  \bibinfo{journal}{Trends Biochem. Sci.} \textbf{\bibinfo{volume}{37}},
  \bibinfo{pages}{526} (\bibinfo{year}{2012}).

\bibitem[{\citenamefont{Simunovic et~al.}(2015)\citenamefont{Simunovic, Voth,
  Callan-Jones, and Bassereau}}]{simu15a}
\bibinfo{author}{\bibfnamefont{M.}~\bibnamefont{Simunovic}},
  \bibinfo{author}{\bibfnamefont{G.~A.} \bibnamefont{Voth}},
  \bibinfo{author}{\bibfnamefont{A.}~\bibnamefont{Callan-Jones}},
  \bibnamefont{and}
  \bibinfo{author}{\bibfnamefont{P.}~\bibnamefont{Bassereau}},
  \bibinfo{journal}{Trends Cell Biol.} \textbf{\bibinfo{volume}{25}},
  \bibinfo{pages}{780} (\bibinfo{year}{2015}).

\bibitem[{\citenamefont{Peter et~al.}(2004)\citenamefont{Peter, Kent, Mills,
  Vallis, Butler, Evans, and McMahon}}]{pete04}
\bibinfo{author}{\bibfnamefont{B.~J.} \bibnamefont{Peter}},
  \bibinfo{author}{\bibfnamefont{H.~M.} \bibnamefont{Kent}},
  \bibinfo{author}{\bibfnamefont{I.~G.} \bibnamefont{Mills}},
  \bibinfo{author}{\bibfnamefont{Y.}~\bibnamefont{Vallis}},
  \bibinfo{author}{\bibfnamefont{P.~J.~G.} \bibnamefont{Butler}},
  \bibinfo{author}{\bibfnamefont{P.~R.} \bibnamefont{Evans}}, \bibnamefont{and}
  \bibinfo{author}{\bibfnamefont{H.~T.} \bibnamefont{McMahon}},
  \bibinfo{journal}{Science} \textbf{\bibinfo{volume}{303}},
  \bibinfo{pages}{495} (\bibinfo{year}{2004}).

\bibitem[{\citenamefont{Mattila et~al.}(2007)\citenamefont{Mattila,
  Pyk{\"a}l{\"a}inen, Saarikangas, Paavilainen, Vihinen, Jokitalo, and
  Lappalainen}}]{matt07}
\bibinfo{author}{\bibfnamefont{P.~K.} \bibnamefont{Mattila}},
  \bibinfo{author}{\bibfnamefont{A.}~\bibnamefont{Pyk{\"a}l{\"a}inen}},
  \bibinfo{author}{\bibfnamefont{J.}~\bibnamefont{Saarikangas}},
  \bibinfo{author}{\bibfnamefont{V.~O.} \bibnamefont{Paavilainen}},
  \bibinfo{author}{\bibfnamefont{H.}~\bibnamefont{Vihinen}},
  \bibinfo{author}{\bibfnamefont{E.}~\bibnamefont{Jokitalo}}, \bibnamefont{and}
  \bibinfo{author}{\bibfnamefont{P.}~\bibnamefont{Lappalainen}},
  \bibinfo{journal}{J. Cell Biol.} \textbf{\bibinfo{volume}{176}},
  \bibinfo{pages}{953} (\bibinfo{year}{2007}).

\bibitem[{\citenamefont{Frost et~al.}(2008)\citenamefont{Frost, Perera, Roux,
  Spasov, Destaing, Egelman, {De Camilli}, and Unger}}]{fros08}
\bibinfo{author}{\bibfnamefont{A.}~\bibnamefont{Frost}},
  \bibinfo{author}{\bibfnamefont{R.}~\bibnamefont{Perera}},
  \bibinfo{author}{\bibfnamefont{A.}~\bibnamefont{Roux}},
  \bibinfo{author}{\bibfnamefont{K.}~\bibnamefont{Spasov}},
  \bibinfo{author}{\bibfnamefont{O.}~\bibnamefont{Destaing}},
  \bibinfo{author}{\bibfnamefont{E.~H.} \bibnamefont{Egelman}},
  \bibinfo{author}{\bibfnamefont{P.}~\bibnamefont{{De Camilli}}},
  \bibnamefont{and} \bibinfo{author}{\bibfnamefont{V.~M.} \bibnamefont{Unger}},
  \bibinfo{journal}{Cell} \textbf{\bibinfo{volume}{132}}, \bibinfo{pages}{807}
  (\bibinfo{year}{2008}).

\bibitem[{\citenamefont{Wang et~al.}(2009)\citenamefont{Wang, Navarro, Peng,
  Molinelli, Goh, Judson, Rajashankarc, and Sondermann}}]{wang09}
\bibinfo{author}{\bibfnamefont{Q.}~\bibnamefont{Wang}},
  \bibinfo{author}{\bibfnamefont{M.~V. A.~S.} \bibnamefont{Navarro}},
  \bibinfo{author}{\bibfnamefont{G.}~\bibnamefont{Peng}},
  \bibinfo{author}{\bibfnamefont{E.}~\bibnamefont{Molinelli}},
  \bibinfo{author}{\bibfnamefont{S.~L.} \bibnamefont{Goh}},
  \bibinfo{author}{\bibfnamefont{B.~L.} \bibnamefont{Judson}},
  \bibinfo{author}{\bibfnamefont{K.~R.} \bibnamefont{Rajashankarc}},
  \bibnamefont{and}
  \bibinfo{author}{\bibfnamefont{H.}~\bibnamefont{Sondermann}},
  \bibinfo{journal}{Proc.\ Natl.\ Acad.\ Sci.\ USA}
  \textbf{\bibinfo{volume}{106}}, \bibinfo{pages}{12700}
  (\bibinfo{year}{2009}).

\bibitem[{\citenamefont{Zhu et~al.}(2012)\citenamefont{Zhu, Das, and
  Baumgart}}]{zhu12}
\bibinfo{author}{\bibfnamefont{C.}~\bibnamefont{Zhu}},
  \bibinfo{author}{\bibfnamefont{S.~L.} \bibnamefont{Das}}, \bibnamefont{and}
  \bibinfo{author}{\bibfnamefont{T.}~\bibnamefont{Baumgart}},
  \bibinfo{journal}{Biophys. J.} \textbf{\bibinfo{volume}{102}},
  \bibinfo{pages}{1837} (\bibinfo{year}{2012}).

\bibitem[{\citenamefont{Tanaka-Takiguchi
  et~al.}(2013)\citenamefont{Tanaka-Takiguchi, Itoh, Tsujita, Yamada,
  Yanagisawa, Fujiwara, Yamamoto, Ichikawa, and Takiguchi}}]{tana13}
\bibinfo{author}{\bibfnamefont{Y.}~\bibnamefont{Tanaka-Takiguchi}},
  \bibinfo{author}{\bibfnamefont{T.}~\bibnamefont{Itoh}},
  \bibinfo{author}{\bibfnamefont{K.}~\bibnamefont{Tsujita}},
  \bibinfo{author}{\bibfnamefont{S.}~\bibnamefont{Yamada}},
  \bibinfo{author}{\bibfnamefont{M.}~\bibnamefont{Yanagisawa}},
  \bibinfo{author}{\bibfnamefont{K.}~\bibnamefont{Fujiwara}},
  \bibinfo{author}{\bibfnamefont{A.}~\bibnamefont{Yamamoto}},
  \bibinfo{author}{\bibfnamefont{M.}~\bibnamefont{Ichikawa}}, \bibnamefont{and}
  \bibinfo{author}{\bibfnamefont{K.}~\bibnamefont{Takiguchi}},
  \bibinfo{journal}{Langmuir} \textbf{\bibinfo{volume}{29}},
  \bibinfo{pages}{328} (\bibinfo{year}{2013}).

\bibitem[{\citenamefont{Shi and Baumbart}(2015)}]{shi15}
\bibinfo{author}{\bibfnamefont{Z.}~\bibnamefont{Shi}} \bibnamefont{and}
  \bibinfo{author}{\bibfnamefont{T.}~\bibnamefont{Baumbart}},
  \bibinfo{journal}{Nature Comm.} \textbf{\bibinfo{volume}{6}},
  \bibinfo{pages}{5974} (\bibinfo{year}{2015}).

\bibitem[{\citenamefont{Pr{\'e}vost et~al.}(2015)\citenamefont{Pr{\'e}vost,
  Zhao, Manzi, Lemichez, Lappalainen, Callan-Jones, and Bassereau}}]{prev15}
\bibinfo{author}{\bibfnamefont{C.}~\bibnamefont{Pr{\'e}vost}},
  \bibinfo{author}{\bibfnamefont{H.}~\bibnamefont{Zhao}},
  \bibinfo{author}{\bibfnamefont{J.}~\bibnamefont{Manzi}},
  \bibinfo{author}{\bibfnamefont{E.}~\bibnamefont{Lemichez}},
  \bibinfo{author}{\bibfnamefont{P.}~\bibnamefont{Lappalainen}},
  \bibinfo{author}{\bibfnamefont{A.}~\bibnamefont{Callan-Jones}},
  \bibnamefont{and}
  \bibinfo{author}{\bibfnamefont{P.}~\bibnamefont{Bassereau}},
  \bibinfo{journal}{Nature Comm.} \textbf{\bibinfo{volume}{6}},
  \bibinfo{pages}{8529} (\bibinfo{year}{2015}).

\bibitem[{\citenamefont{Isas et~al.}(2015)\citenamefont{Isas, Ambroso, Hegde,
  Langen, and Langen}}]{isas15}
\bibinfo{author}{\bibfnamefont{J.~M.} \bibnamefont{Isas}},
  \bibinfo{author}{\bibfnamefont{M.~R.} \bibnamefont{Ambroso}},
  \bibinfo{author}{\bibfnamefont{P.~B.} \bibnamefont{Hegde}},
  \bibinfo{author}{\bibfnamefont{J.}~\bibnamefont{Langen}}, \bibnamefont{and}
  \bibinfo{author}{\bibfnamefont{R.}~\bibnamefont{Langen}},
  \bibinfo{journal}{Structure} \textbf{\bibinfo{volume}{23}},
  \bibinfo{pages}{873} (\bibinfo{year}{2015}).

\bibitem[{\citenamefont{Adam et~al.}(2015)\citenamefont{Adam, Basnet, and
  Mizuno}}]{adam15}
\bibinfo{author}{\bibfnamefont{J.}~\bibnamefont{Adam}},
  \bibinfo{author}{\bibfnamefont{N.}~\bibnamefont{Basnet}}, \bibnamefont{and}
  \bibinfo{author}{\bibfnamefont{N.}~\bibnamefont{Mizuno}},
  \bibinfo{journal}{Sci. Rep.} \textbf{\bibinfo{volume}{5}},
  \bibinfo{pages}{15452} (\bibinfo{year}{2015}).

\bibitem[{\citenamefont{Lipowsky}(2013)}]{lipo13}
\bibinfo{author}{\bibfnamefont{R.}~\bibnamefont{Lipowsky}},
  \bibinfo{journal}{Faraday Discuss.} \textbf{\bibinfo{volume}{161}},
  \bibinfo{pages}{305} (\bibinfo{year}{2013}).

\bibitem[{\citenamefont{G{\'o}{\'z}d{\'z}}(2006)}]{gozd06}
\bibinfo{author}{\bibfnamefont{W.~T.} \bibnamefont{G{\'o}{\'z}d{\'z}}},
  \bibinfo{journal}{J. Phys. Chem. B} \textbf{\bibinfo{volume}{110}},
  \bibinfo{pages}{21981} (\bibinfo{year}{2006}).

\bibitem[{\citenamefont{Phillips et~al.}(2009)\citenamefont{Phillips, Ursell,
  Wiggins, and Sens}}]{phil09}
\bibinfo{author}{\bibfnamefont{R.}~\bibnamefont{Phillips}},
  \bibinfo{author}{\bibfnamefont{T.}~\bibnamefont{Ursell}},
  \bibinfo{author}{\bibfnamefont{P.}~\bibnamefont{Wiggins}}, \bibnamefont{and}
  \bibinfo{author}{\bibfnamefont{P.}~\bibnamefont{Sens}},
  \bibinfo{journal}{Nature} \textbf{\bibinfo{volume}{459}},
  \bibinfo{pages}{379} (\bibinfo{year}{2009}).

\bibitem[{\citenamefont{Greenall and Gompper}(2011)}]{gree11}
\bibinfo{author}{\bibfnamefont{M.~J.} \bibnamefont{Greenall}} \bibnamefont{and}
  \bibinfo{author}{\bibfnamefont{G.}~\bibnamefont{Gompper}},
  \bibinfo{journal}{Langmuir} \textbf{\bibinfo{volume}{27}},
  \bibinfo{pages}{3416} (\bibinfo{year}{2011}).

\bibitem[{\citenamefont{Noguchi}(2012)}]{nogu12a}
\bibinfo{author}{\bibfnamefont{H.}~\bibnamefont{Noguchi}},
  \bibinfo{journal}{Soft Matter} \textbf{\bibinfo{volume}{8}},
  \bibinfo{pages}{8926} (\bibinfo{year}{2012}).

\bibitem[{\citenamefont{Aimon et~al.}(2014)\citenamefont{Aimon, Callan-Jones,
  Berthaud, Pinot, Toombes, and Bassereau}}]{aimo14}
\bibinfo{author}{\bibfnamefont{S.}~\bibnamefont{Aimon}},
  \bibinfo{author}{\bibfnamefont{A.}~\bibnamefont{Callan-Jones}},
  \bibinfo{author}{\bibfnamefont{A.}~\bibnamefont{Berthaud}},
  \bibinfo{author}{\bibfnamefont{M.}~\bibnamefont{Pinot}},
  \bibinfo{author}{\bibfnamefont{G.~E.} \bibnamefont{Toombes}},
  \bibnamefont{and}
  \bibinfo{author}{\bibfnamefont{P.}~\bibnamefont{Bassereau}},
  \bibinfo{journal}{Dev. Cell} \textbf{\bibinfo{volume}{28}},
  \bibinfo{pages}{212} (\bibinfo{year}{2014}).

\bibitem[{\citenamefont{Reynwar et~al.}(2007)\citenamefont{Reynwar, Ilya,
  Harmandaris, M{\"u}ller, Kremer, and Deserno}}]{reyn07}
\bibinfo{author}{\bibfnamefont{B.~J.} \bibnamefont{Reynwar}},
  \bibinfo{author}{\bibfnamefont{G.}~\bibnamefont{Ilya}},
  \bibinfo{author}{\bibfnamefont{V.~A.} \bibnamefont{Harmandaris}},
  \bibinfo{author}{\bibfnamefont{M.~M.} \bibnamefont{M{\"u}ller}},
  \bibinfo{author}{\bibfnamefont{K.}~\bibnamefont{Kremer}}, \bibnamefont{and}
  \bibinfo{author}{\bibfnamefont{M.}~\bibnamefont{Deserno}},
  \bibinfo{journal}{Nature} \textbf{\bibinfo{volume}{447}},
  \bibinfo{pages}{461} (\bibinfo{year}{2007}).

\bibitem[{\citenamefont{Auth and Gompper}(2009)}]{auth09}
\bibinfo{author}{\bibfnamefont{T.}~\bibnamefont{Auth}} \bibnamefont{and}
  \bibinfo{author}{\bibfnamefont{G.}~\bibnamefont{Gompper}},
  \bibinfo{journal}{Phys. Rev. E} \textbf{\bibinfo{volume}{80}},
  \bibinfo{pages}{031901} (\bibinfo{year}{2009}).

\bibitem[{\citenamefont{{\v S}ari{\'c} and Cacciuto}(2012)}]{sari12}
\bibinfo{author}{\bibfnamefont{A.}~\bibnamefont{{\v S}ari{\'c}}}
  \bibnamefont{and} \bibinfo{author}{\bibfnamefont{A.}~\bibnamefont{Cacciuto}},
  \bibinfo{journal}{Phys. Rev. Lett.} \textbf{\bibinfo{volume}{108}},
  \bibinfo{pages}{118101} (\bibinfo{year}{2012}).

\bibitem[{\citenamefont{G{\'o}mez-Llobregat
  et~al.}(2016)\citenamefont{G{\'o}mez-Llobregat, El{\'i}as-Wolff, and
  Lind{\'e}n}}]{gome16}
\bibinfo{author}{\bibfnamefont{J.}~\bibnamefont{G{\'o}mez-Llobregat}},
  \bibinfo{author}{\bibfnamefont{F.}~\bibnamefont{El{\'i}as-Wolff}},
  \bibnamefont{and}
  \bibinfo{author}{\bibfnamefont{M.}~\bibnamefont{Lind{\'e}n}},
  \bibinfo{journal}{Biophys. J.} \textbf{\bibinfo{volume}{110}},
  \bibinfo{pages}{197} (\bibinfo{year}{2016}).

\bibitem[{\citenamefont{Canham}(1970)}]{canh70}
\bibinfo{author}{\bibfnamefont{P.~B.} \bibnamefont{Canham}},
  \bibinfo{journal}{J. Theor. Biol.} \textbf{\bibinfo{volume}{26}},
  \bibinfo{pages}{61} (\bibinfo{year}{1970}).

\bibitem[{\citenamefont{Helfrich}(1973)}]{helf73}
\bibinfo{author}{\bibfnamefont{W.}~\bibnamefont{Helfrich}},
  \bibinfo{journal}{Z.\ Naturforsch} \textbf{\bibinfo{volume}{28c}},
  \bibinfo{pages}{693} (\bibinfo{year}{1973}).

\bibitem[{\citenamefont{Fournier}(1996)}]{four96}
\bibinfo{author}{\bibfnamefont{J.-B.} \bibnamefont{Fournier}},
  \bibinfo{journal}{Phys. Rev. Lett.} \textbf{\bibinfo{volume}{76}},
  \bibinfo{pages}{4436} (\bibinfo{year}{1996}).

\bibitem[{\citenamefont{Kabaso et~al.}(2011)\citenamefont{Kabaso, Gongadze,
  Elter, van Rienen, Gimsa, Kralj-Igli{\v{c}}, and Igli{\v{c}}}}]{kaba11}
\bibinfo{author}{\bibfnamefont{D.}~\bibnamefont{Kabaso}},
  \bibinfo{author}{\bibfnamefont{E.}~\bibnamefont{Gongadze}},
  \bibinfo{author}{\bibfnamefont{P.}~\bibnamefont{Elter}},
  \bibinfo{author}{\bibfnamefont{U.}~\bibnamefont{van Rienen}},
  \bibinfo{author}{\bibfnamefont{J.}~\bibnamefont{Gimsa}},
  \bibinfo{author}{\bibfnamefont{V.}~\bibnamefont{Kralj-Igli{\v{c}}}},
  \bibnamefont{and}
  \bibinfo{author}{\bibfnamefont{A.}~\bibnamefont{Igli{\v{c}}}},
  \bibinfo{journal}{Mini Rev. Med. Chem.} \textbf{\bibinfo{volume}{11}},
  \bibinfo{pages}{272} (\bibinfo{year}{2011}).

\bibitem[{\citenamefont{Igli{\v{c}} et~al.}(2006)\citenamefont{Igli{\v{c}},
  H{\"{a}}gerstrand, Verani{\v{c}}, Plemenita{\v{s}}, and
  Kralj-Igli{\v{c}}}}]{igli06}
\bibinfo{author}{\bibfnamefont{A.}~\bibnamefont{Igli{\v{c}}}},
  \bibinfo{author}{\bibfnamefont{H.}~\bibnamefont{H{\"{a}}gerstrand}},
  \bibinfo{author}{\bibfnamefont{P.}~\bibnamefont{Verani{\v{c}}}},
  \bibinfo{author}{\bibfnamefont{A.}~\bibnamefont{Plemenita{\v{s}}}},
  \bibnamefont{and}
  \bibinfo{author}{\bibfnamefont{V.}~\bibnamefont{Kralj-Igli{\v{c}}}},
  \bibinfo{journal}{J. Theor. Biol.} \textbf{\bibinfo{volume}{240}},
  \bibinfo{pages}{368} (\bibinfo{year}{2006}).

\bibitem[{\citenamefont{Dommersnes and Fournier}(1999)}]{domm99}
\bibinfo{author}{\bibfnamefont{P.~G.} \bibnamefont{Dommersnes}}
  \bibnamefont{and} \bibinfo{author}{\bibfnamefont{J.~B.}
  \bibnamefont{Fournier}}, \bibinfo{journal}{Eur. Phys. J. B}
  \textbf{\bibinfo{volume}{12}}, \bibinfo{pages}{9} (\bibinfo{year}{1999}).

\bibitem[{\citenamefont{Dommersnes and Fournier}(2002)}]{domm02}
\bibinfo{author}{\bibfnamefont{P.~G.} \bibnamefont{Dommersnes}}
  \bibnamefont{and} \bibinfo{author}{\bibfnamefont{J.~B.}
  \bibnamefont{Fournier}}, \bibinfo{journal}{Biophys. J.}
  \textbf{\bibinfo{volume}{83}}, \bibinfo{pages}{2898} (\bibinfo{year}{2002}).

\bibitem[{\citenamefont{Schweitzer and Kozlov}(2015)}]{schw15}
\bibinfo{author}{\bibfnamefont{Y.}~\bibnamefont{Schweitzer}} \bibnamefont{and}
  \bibinfo{author}{\bibfnamefont{M.~M.} \bibnamefont{Kozlov}},
  \bibinfo{journal}{PLoS Comput. Biol.} \textbf{\bibinfo{volume}{11}},
  \bibinfo{pages}{e1004054} (\bibinfo{year}{2015}).

\bibitem[{\citenamefont{Blood and Voth}(2006)}]{bloo06}
\bibinfo{author}{\bibfnamefont{P.~D.} \bibnamefont{Blood}} \bibnamefont{and}
  \bibinfo{author}{\bibfnamefont{G.~A.} \bibnamefont{Voth}},
  \bibinfo{journal}{Proc.\ Natl.\ Acad.\ Sci.\ USA}
  \textbf{\bibinfo{volume}{103}}, \bibinfo{pages}{15068}
  (\bibinfo{year}{2006}).

\bibitem[{\citenamefont{Arkhipov et~al.}(2008)\citenamefont{Arkhipov, Yin, and
  Schulten}}]{arkh08}
\bibinfo{author}{\bibfnamefont{A.}~\bibnamefont{Arkhipov}},
  \bibinfo{author}{\bibfnamefont{Y.}~\bibnamefont{Yin}}, \bibnamefont{and}
  \bibinfo{author}{\bibfnamefont{K.}~\bibnamefont{Schulten}},
  \bibinfo{journal}{Biophys.\ J.} \textbf{\bibinfo{volume}{95}},
  \bibinfo{pages}{2806} (\bibinfo{year}{2008}).

\bibitem[{\citenamefont{Khelashvili et~al.}(2009)\citenamefont{Khelashvili,
  Harries, and Weinstein}}]{khel09}
\bibinfo{author}{\bibfnamefont{G.}~\bibnamefont{Khelashvili}},
  \bibinfo{author}{\bibfnamefont{D.}~\bibnamefont{Harries}}, \bibnamefont{and}
  \bibinfo{author}{\bibfnamefont{H.}~\bibnamefont{Weinstein}},
  \bibinfo{journal}{Biophys. J.} \textbf{\bibinfo{volume}{97}},
  \bibinfo{pages}{1626} (\bibinfo{year}{2009}).

\bibitem[{\citenamefont{Yu and Schulten}(2013)}]{yu13}
\bibinfo{author}{\bibfnamefont{H.}~\bibnamefont{Yu}} \bibnamefont{and}
  \bibinfo{author}{\bibfnamefont{K.}~\bibnamefont{Schulten}},
  \bibinfo{journal}{PLoS Comput. Biol.} \textbf{\bibinfo{volume}{9}},
  \bibinfo{pages}{e1002892} (\bibinfo{year}{2013}).

\bibitem[{\citenamefont{Simunovic
  et~al.}(2013{\natexlab{a}})\citenamefont{Simunovic, Srivastava, and
  Voth}}]{simu13}
\bibinfo{author}{\bibfnamefont{M.}~\bibnamefont{Simunovic}},
  \bibinfo{author}{\bibfnamefont{A.}~\bibnamefont{Srivastava}},
  \bibnamefont{and} \bibinfo{author}{\bibfnamefont{G.~A.} \bibnamefont{Voth}},
  \bibinfo{journal}{Proc.\ Natl.\ Acad.\ Sci.\ USA}
  \textbf{\bibinfo{volume}{110}}, \bibinfo{pages}{20396}
  (\bibinfo{year}{2013}{\natexlab{a}}).

\bibitem[{\citenamefont{Simunovic and Voth}(2015)}]{simu15}
\bibinfo{author}{\bibfnamefont{M.}~\bibnamefont{Simunovic}} \bibnamefont{and}
  \bibinfo{author}{\bibfnamefont{G.~A.} \bibnamefont{Voth}},
  \bibinfo{journal}{Nature Comm.} \textbf{\bibinfo{volume}{6}},
  \bibinfo{pages}{7219} (\bibinfo{year}{2015}).

\bibitem[{\citenamefont{Ramakrishnan et~al.}(2012)\citenamefont{Ramakrishnan,
  Ipsen, and {Sunil Kumar}}}]{rama12}
\bibinfo{author}{\bibfnamefont{N.}~\bibnamefont{Ramakrishnan}},
  \bibinfo{author}{\bibfnamefont{J.~H.} \bibnamefont{Ipsen}}, \bibnamefont{and}
  \bibinfo{author}{\bibfnamefont{P.~B.} \bibnamefont{{Sunil Kumar}}},
  \bibinfo{journal}{Soft Matter} \textbf{\bibinfo{volume}{8}},
  \bibinfo{pages}{3058} (\bibinfo{year}{2012}).

\bibitem[{\citenamefont{Ramakrishnan et~al.}(2013)\citenamefont{Ramakrishnan,
  {Sunil Kumar}, and Ipsen}}]{rama13}
\bibinfo{author}{\bibfnamefont{N.}~\bibnamefont{Ramakrishnan}},
  \bibinfo{author}{\bibfnamefont{P.~B.} \bibnamefont{{Sunil Kumar}}},
  \bibnamefont{and} \bibinfo{author}{\bibfnamefont{J.~H.} \bibnamefont{Ipsen}},
  \bibinfo{journal}{Biophys. J.} \textbf{\bibinfo{volume}{104}},
  \bibinfo{pages}{1018} (\bibinfo{year}{2013}).

\bibitem[{\citenamefont{Noguchi}(2014)}]{nogu14}
\bibinfo{author}{\bibfnamefont{H.}~\bibnamefont{Noguchi}},
  \bibinfo{journal}{EPL} \textbf{\bibinfo{volume}{108}}, \bibinfo{pages}{48001}
  (\bibinfo{year}{2014}).

\bibitem[{\citenamefont{Noguchi}(2015)}]{nogu15b}
\bibinfo{author}{\bibfnamefont{H.}~\bibnamefont{Noguchi}}, \bibinfo{journal}{J.
  Chem. Phys.} \textbf{\bibinfo{volume}{143}}, \bibinfo{pages}{243109}
  (\bibinfo{year}{2015}).

\bibitem[{\citenamefont{Noguchi}(2016)}]{nogu16}
\bibinfo{author}{\bibfnamefont{H.}~\bibnamefont{Noguchi}},
  \bibinfo{journal}{Sci.\ Rep.} \textbf{\bibinfo{volume}{6}},
  \bibinfo{pages}{20935} (\bibinfo{year}{2016}).

\bibitem[{\citenamefont{Ayton et~al.}(2009)\citenamefont{Ayton, Lyman, Krishna,
  Swenson, Mim, Unger, and Voth}}]{ayto09}
\bibinfo{author}{\bibfnamefont{G.~S.} \bibnamefont{Ayton}},
  \bibinfo{author}{\bibfnamefont{E.}~\bibnamefont{Lyman}},
  \bibinfo{author}{\bibfnamefont{V.}~\bibnamefont{Krishna}},
  \bibinfo{author}{\bibfnamefont{R.~D.} \bibnamefont{Swenson}},
  \bibinfo{author}{\bibfnamefont{C.}~\bibnamefont{Mim}},
  \bibinfo{author}{\bibfnamefont{V.~M.} \bibnamefont{Unger}}, \bibnamefont{and}
  \bibinfo{author}{\bibfnamefont{G.~A.} \bibnamefont{Voth}},
  \bibinfo{journal}{Biophys. J.} \textbf{\bibinfo{volume}{97}},
  \bibinfo{pages}{1616} (\bibinfo{year}{2009}).

\bibitem[{\citenamefont{Simunovic
  et~al.}(2013{\natexlab{b}})\citenamefont{Simunovic, Mim, Marlovits, Resch,
  Unger, and Voth}}]{simu13a}
\bibinfo{author}{\bibfnamefont{M.}~\bibnamefont{Simunovic}},
  \bibinfo{author}{\bibfnamefont{C.}~\bibnamefont{Mim}},
  \bibinfo{author}{\bibfnamefont{T.~C.} \bibnamefont{Marlovits}},
  \bibinfo{author}{\bibfnamefont{G.}~\bibnamefont{Resch}},
  \bibinfo{author}{\bibfnamefont{V.~M.} \bibnamefont{Unger}}, \bibnamefont{and}
  \bibinfo{author}{\bibfnamefont{G.~A.} \bibnamefont{Voth}},
  \bibinfo{journal}{Biophys. J.} \textbf{\bibinfo{volume}{105}},
  \bibinfo{pages}{711} (\bibinfo{year}{2013}{\natexlab{b}}).

\bibitem[{\citenamefont{Noguchi}(2009)}]{nogu09}
\bibinfo{author}{\bibfnamefont{H.}~\bibnamefont{Noguchi}},
  \bibinfo{journal}{J.\ Phys.\ Soc.\ Jpn.} \textbf{\bibinfo{volume}{78}},
  \bibinfo{pages}{041007} (\bibinfo{year}{2009}).

\bibitem[{\citenamefont{Noguchi and Gompper}(2006)}]{nogu06}
\bibinfo{author}{\bibfnamefont{H.}~\bibnamefont{Noguchi}} \bibnamefont{and}
  \bibinfo{author}{\bibfnamefont{G.}~\bibnamefont{Gompper}},
  \bibinfo{journal}{Phys.\ Rev.\ E} \textbf{\bibinfo{volume}{73}},
  \bibinfo{pages}{021903} (\bibinfo{year}{2006}).

\bibitem[{\citenamefont{Shiba and Noguchi}(2011)}]{shib11}
\bibinfo{author}{\bibfnamefont{H.}~\bibnamefont{Shiba}} \bibnamefont{and}
  \bibinfo{author}{\bibfnamefont{H.}~\bibnamefont{Noguchi}},
  \bibinfo{journal}{Phys. Rev. E} \textbf{\bibinfo{volume}{84}},
  \bibinfo{pages}{031926} (\bibinfo{year}{2011}).

\bibitem[{\citenamefont{Noguchi}(2013)}]{nogu13}
\bibinfo{author}{\bibfnamefont{H.}~\bibnamefont{Noguchi}},
  \bibinfo{journal}{EPL} \textbf{\bibinfo{volume}{102}}, \bibinfo{pages}{68001}
  (\bibinfo{year}{2013}).

\bibitem[{\citenamefont{Allen and Tildesley}(1987)}]{alle87}
\bibinfo{author}{\bibfnamefont{M.~P.} \bibnamefont{Allen}} \bibnamefont{and}
  \bibinfo{author}{\bibfnamefont{D.~J.} \bibnamefont{Tildesley}},
  \emph{\bibinfo{title}{Computer Simulation of Liquids}}
  (\bibinfo{publisher}{Clarendon Press}, \bibinfo{address}{Oxford},
  \bibinfo{year}{1987}).

\bibitem[{\citenamefont{Noguchi}(2011)}]{nogu11}
\bibinfo{author}{\bibfnamefont{H.}~\bibnamefont{Noguchi}},
  \bibinfo{journal}{J.\ Chem.\ Phys.} \textbf{\bibinfo{volume}{134}},
  \bibinfo{pages}{055101} (\bibinfo{year}{2011}).

\bibitem[{\citenamefont{Ramadurai et~al.}(2009)\citenamefont{Ramadurai, Holt,
  Krasnikov, van~den Bogaart, Killian, and Poolman}}]{rama09a}
\bibinfo{author}{\bibfnamefont{S.}~\bibnamefont{Ramadurai}},
  \bibinfo{author}{\bibfnamefont{A.}~\bibnamefont{Holt}},
  \bibinfo{author}{\bibfnamefont{V.}~\bibnamefont{Krasnikov}},
  \bibinfo{author}{\bibfnamefont{G.}~\bibnamefont{van~den Bogaart}},
  \bibinfo{author}{\bibfnamefont{J.~A.} \bibnamefont{Killian}},
  \bibnamefont{and} \bibinfo{author}{\bibfnamefont{B.}~\bibnamefont{Poolman}},
  \bibinfo{journal}{J. Am. Chem. Soc.} \textbf{\bibinfo{volume}{131}},
  \bibinfo{pages}{12650} (\bibinfo{year}{2009}).

\bibitem[{\citenamefont{Hukushima and Nemoto}(1996)}]{huku96}
\bibinfo{author}{\bibfnamefont{K.}~\bibnamefont{Hukushima}} \bibnamefont{and}
  \bibinfo{author}{\bibfnamefont{K.}~\bibnamefont{Nemoto}},
  \bibinfo{journal}{J.\ Phys.\ Soc.\ Jpn.} \textbf{\bibinfo{volume}{65}},
  \bibinfo{pages}{1604} (\bibinfo{year}{1996}).

\bibitem[{\citenamefont{Okamoto}(2004)}]{okam04}
\bibinfo{author}{\bibfnamefont{Y.}~\bibnamefont{Okamoto}}, \bibinfo{journal}{J.
  Mol. Graph. Model.} \textbf{\bibinfo{volume}{22}}, \bibinfo{pages}{425}
  (\bibinfo{year}{2004}).

\bibitem[{\citenamefont{Wu et~al.}(2013)\citenamefont{Wu, Shiba, and
  Noguchi}}]{wu13}
\bibinfo{author}{\bibfnamefont{H.}~\bibnamefont{Wu}},
  \bibinfo{author}{\bibfnamefont{H.}~\bibnamefont{Shiba}}, \bibnamefont{and}
  \bibinfo{author}{\bibfnamefont{H.}~\bibnamefont{Noguchi}},
  \bibinfo{journal}{Soft matter} \textbf{\bibinfo{volume}{9}},
  \bibinfo{pages}{9907} (\bibinfo{year}{2013}).

\bibitem[{\citenamefont{Tolpekina et~al.}(2004)\citenamefont{Tolpekina, den
  Otter, and Briels}}]{tolp04}
\bibinfo{author}{\bibfnamefont{T.~V.} \bibnamefont{Tolpekina}},
  \bibinfo{author}{\bibfnamefont{W.~K.} \bibnamefont{den Otter}},
  \bibnamefont{and} \bibinfo{author}{\bibfnamefont{W.~J.}
  \bibnamefont{Briels}}, \bibinfo{journal}{J.\ Chem.\ Phys.}
  \textbf{\bibinfo{volume}{121}}, \bibinfo{pages}{8014} (\bibinfo{year}{2004}).

\bibitem[{\citenamefont{Fromherz}(1983)}]{from83}
\bibinfo{author}{\bibfnamefont{P.}~\bibnamefont{Fromherz}},
  \bibinfo{journal}{Chem.\ Phys.\ Lett.} \textbf{\bibinfo{volume}{94}},
  \bibinfo{pages}{259} (\bibinfo{year}{1983}).

\bibitem[{\citenamefont{Hu et~al.}(2012)\citenamefont{Hu, Briguglio, and
  Deserno}}]{hu12}
\bibinfo{author}{\bibfnamefont{M.}~\bibnamefont{Hu}},
  \bibinfo{author}{\bibfnamefont{J.~J.} \bibnamefont{Briguglio}},
  \bibnamefont{and} \bibinfo{author}{\bibfnamefont{M.}~\bibnamefont{Deserno}},
  \bibinfo{journal}{Biophys. J.} \textbf{\bibinfo{volume}{102}},
  \bibinfo{pages}{1403} (\bibinfo{year}{2012}).

\bibitem[{\citenamefont{Nakagawa and Noguchi}(2015)}]{naka15}
\bibinfo{author}{\bibfnamefont{K.~M.} \bibnamefont{Nakagawa}} \bibnamefont{and}
  \bibinfo{author}{\bibfnamefont{H.}~\bibnamefont{Noguchi}},
  \bibinfo{journal}{Soft matter} \textbf{\bibinfo{volume}{11}},
  \bibinfo{pages}{1403} (\bibinfo{year}{2015}).

\bibitem[{\citenamefont{Nomura et~al.}(2001)\citenamefont{Nomura, Nagata,
  Inaba, Hiramatsu, Hotani, and Takiguchi}}]{nomu01}
\bibinfo{author}{\bibfnamefont{F.}~\bibnamefont{Nomura}},
  \bibinfo{author}{\bibfnamefont{M.}~\bibnamefont{Nagata}},
  \bibinfo{author}{\bibfnamefont{T.}~\bibnamefont{Inaba}},
  \bibinfo{author}{\bibfnamefont{H.}~\bibnamefont{Hiramatsu}},
  \bibinfo{author}{\bibfnamefont{H.}~\bibnamefont{Hotani}}, \bibnamefont{and}
  \bibinfo{author}{\bibfnamefont{K.}~\bibnamefont{Takiguchi}},
  \bibinfo{journal}{Proc.\ Natl.\ Acad.\ Sci.\ USA}
  \textbf{\bibinfo{volume}{98}}, \bibinfo{pages}{2340} (\bibinfo{year}{2001}).

\end{thebibliography}

\end{document}